\documentclass[12pt]{iopart}
\usepackage{iopams}
\usepackage{xcolor}
\usepackage{colortbl}
\usepackage{pstricks,multido}
\usepackage{wrapfig}
\usepackage{graphicx,epsfig}
\usepackage[margin=2cm,top=2.5cm,bottom=2.3cm,centering]{geometry}

\definecolor{lightyellow}{rgb}{.95,.96,.5}
\definecolor{myblue}{rgb}{.39,.39,0.92}
\definecolor{midblue}{rgb}{.7,.7,1}
\definecolor{lightblue}{rgb}{.8,.8,1}
\definecolor{macblue}{rgb}{.40,.55,0.84}
\definecolor{mygrey}{rgb}{.75,.75,.75}
\definecolor{myred}{rgb}{1,0.61,0.61}
\definecolor{lightred}{rgb}{0.82,0.23,0.13}
\definecolor{mygreen}{rgb}{0.40,0.75,0.16}
\definecolor{mybrown}{rgb}{0.69,0.49,0.30}

\usepackage{enumitem}

\catcode`\@=11
\def\numberbysection{\@addtoreset{equation}{section}
\def\theequation{\arabic{section}.\arabic{equation}}}
\numberbysection

\def\be{\begin{equation}}
\def\ee{\end{equation}}
\newcommand\bea{\begin{eqnarray}}
\newcommand\eea{\end{eqnarray}}
\renewcommand\phi{\varphi}

\newcommand\egal{&\!\!\!=\!\!\!&}

\def\benn{\begin{eqnarray*}}
\def\eenn{\end{eqnarray*}}

\def\la{\langle}
\def\ra{\rangle}
\def\Z{{\mathbb Z}}

\def\C{{\mathbb C}}

\def\L{{\cal L}}

\def\a{\alpha}
\def\b{\beta}

\def\s{\sigma}
\def\bi{{\mathbb I}}

\def\hdimer{\pspolygon[linearc=.35,linecolor=black,fillstyle=solid,fillcolor=lightgray](-0.35,-0.35)(-0.35,0.35)(1.35,0.35)(1.35,-0.35)}
\def\vdimer{\pspolygon[linearc=.35,linecolor=black,fillstyle=solid,fillcolor=mybrown](-0.35,-0.35)(-0.35,1.35)(0.35,1.35)(0.35,-0.35)}

\def\aup{\psline[linewidth=2pt,linecolor=mybrown]{->}(0,0.10)(0,0.95)}
\def\ado{\psline[linewidth=2pt,linecolor=gray]{<-}(0,0.05)(0,0.90)}
\begin{document}
\hfill\today

\title[]{\Large Transfer matrix for spanning trees, webs and colored forests}

\author{J.G. Brankov$^{1,2}$, V.S. Poghosyan$^3$, V.B. Priezzhev$^1$, P. Ruelle$^4$}

\address{$^1$ Bogoliubov Laboratory of Theoretical Physics, Joint Institute for Nuclear Research, 141980 Dubna, Russia\\
$^2$ Institute of Mechanics, Bulgarian Academy of Sciences, 1113 Sofia, Bulgaria\\
$^3$ Institute for Informatics and Automation Problems NAS of Armenia, 375044 Yerevan, Armenia\\
$^4$ Institute for Research in Mathematics and Physics, Universit\'e catholique de Louvain, B-1348 Louvain-la-Neuve, Belgium}
\begin{abstract}
We use the transfer matrix formalism for dimers proposed by Lieb, and generalize it to address the corresponding problem for arrow configurations (or trees) associated to dimer configurations through Temperley's correspondence. On a cylinder, the arrow configurations can be partitioned into sectors according to the number of non-contractible loops they contain. We show how Lieb's transfer matrix can be adapted in order to disentangle the various sectors and to compute the corresponding partition functions. In order to address the issue of Jordan cells, we introduce a new, extended transfer matrix, which not only keeps track of the positions of the dimers, but also propagates colors along the branches of the associated trees. We argue that this new matrix contains Jordan cells.
\end{abstract}



\section{Introduction}

Spanning trees are a classical theme which forms a chapter of virtually every textbook of combinatorics or graph theory, see for instance \cite{stanley}. For $G$ a connected, unoriented graph, a spanning tree on $G$ is a connected subgraph with the same set of nodes as $G$, and which contains no cycle. The problem of counting the spanning trees on a given graph appears to have a classical solution as well, based on Kirchhoff's theorem (or the Matrix-Tree theorem) \cite{kir}. 

Spanning trees are related to various statistical physics models, like the Potts model \cite{wu}, the dimer model \cite{temp,kpw,ks} and the two-dimensional Abelian sandpile model \cite{dharmaj} in which they play a crucial role. Spanning trees have also been intensively studied in the mathematical literature, where many of the most recent contributions are based on the relation between spanning trees and the loop-erased random walk (LERW) \cite{law}: the chemical paths connecting two points on uniform spanning trees have the same distribution as the LERW running between the same two points \cite{bro,pem,maj,wil}.

As introduced above, a spanning tree is an unoriented structure. By selecting a specific vertex as a root, the spanning tree acquires a natural orientation: the edges forming the spanning tree become oriented and are directed towards the root. A rooted spanning forest is an oriented spanning graph which is a disjoint union of rooted trees. These oriented structures can be seen as configurations of arrows: at each vertex, except the root, there is one and only one outgoing arrow, pointing to one of its neighbouring vertices, potentially the root as the case may be. A rooted spanning tree or forest is then a configuration of arrows which contain no loop, with all arrows flowing to the root. A colored forest is a forest in which the different disconnected trees are assigned distinct colors. 

Relaxing the constraint that the arrow configurations cannot contain loops leads to generalized structures. They are known in the mathematical literature as unicycles, cycle-rooted spanning forests, cycle-rooted spanning groves, which in general combine trees and loops, contractible or not, oriented or not, see for instance \cite{bupe,kenyon2000_1,kenyon2000_2,lsw,kenyon2011,lepe,kenwil2011,kenyon2012} for recent works. In the physical literature, these different structures are known under the general name of spanning webs, and appear most often in the context of the monomer-dimer model \cite{bbgj,ppr1,bgpt,ppr2,popr}.

Our purpose in this article is to study the statistics of spanning webs on two-dimensional square grids. We recall that Temperley's correspondence establishes an isomorphism between the dimer configurations on a rectangular grid and spanning trees or spanning forests on an appropriate sublattice. Depending on the parities of the dimensions of the dimer grid, different boundary conditions may be achieved through the existence or not of connections between boundary sites and the root. A uniform measure on dimer configurations induces a uniform measure on the corresponding class of spanning forests. 

When the grid on which the dimers live is embedded on a cylinder, Temperley's correspondence gives rise to a special class of rooted spanning webs, which, following \cite{kenyon2011}, we will call {\it incompressible spanning webs}, meaning that their connected components are either rooted spanning trees or cycle-rooted spanning trees where all loops must be non-contractible and self-avoiding. Again uniform dimer configurations lead to uniform incompressible spanning webs. Our aim is to study the statistics of the subclasses of those spanning webs that contain a fixed number of loops, by computing the corresponding partial partition functions. In this sense, the present work is a continuation of \cite{bgpt}.

In Section 2, we recall the Temperley correspondence for dimer configurations on a cylinder and the type of boundary conditions which the associated spanning webs are subjected to. In this way the problem of computing the partition funcions for spanning webs reduces to computing the partition function for dimers. Section 3 indicates how this can be done by using the transfer matrix formalism, following an old proposal by Lieb \cite{lieb}. In Section 4, we show how the subclasses of spanning webs comprising a fixed number of loops can be separated, by assigning appropriate weights to certain bonds on the dimer grid. Section 5 indicates how the proper weighing can be implemented on the transfer matrix, and computes the relevant partial partition functions, for incompressible spanning webs on a cylinder and in the conformal limit. Sections 6 and 7 are devoted to the conformal bulk generating functions. Section 8 introduces colors on disconnected rooted trees and discusses the emergence of Jordan cells.

\section{Spanning webs from dimers}

There is a well-known correspondence between rooted spanning forests on a domain and dimer configurations on a related domain, first observed by Temperley \cite{temp} in the case of a finite portion of $\Z^2$ and generalized, much more recently, in various directions \cite{bupe,kpw,ks}. On a non-planar graph, this correspondence establishes a relation between dimer configurations and incompressible spanning webs. Here we recall this relation in the special case we want to study, namely a cylindrical grid.

We consider a $M \times N$ rectangular grid on $\Z^2$ on which we impose a periodic boundary condition horizontally. The sites of this grid $\L$ are labelled by two integer numbers, a horizontal coordinate between 1 and $N$ and a vertical coordinate between 1 and $M$. We also define $\L_{\rm odd}$ and $\L_{\rm even}$ as the subsets of sites with their coordinates either both odd or both even respectively, the union $\L_{\rm odd} \cup \L_{\rm even}$ being (roughly) half the full grid $\L$. Seen as a (non-planar) graph, the sites of $\L$ form the set of vertices while the edges are those inherited from the lattice $\Z^2$.

A dimer configuration on $\L$ is a subset of edges such that every site of $\L$ belongs to exactly one edge. Equivalently one thinks of placing dimers, little rectangles covering exactly two neighbouring sites, in such a way that every site is covered by exactly one dimer (obviously this is only possible if the number of sites $MN$ is even). By defining a probability measure on the set of dimer configurations, a statistical analysis of their general characteristics can be made. A specific (unnormalized) measure which is commonly used is the one that gives a dimer configuration the statistical weight $\alpha ^h \beta^v$, where $\alpha,\beta$ are positive real numbers and $h$ and $v$ are respectively the number of horizontal and vertical dimers contained in the configuration. In what follows, we will use another measure in order to disentangle the various classes of spanning webs to which the dimer configurations will be associated. 

\begin{figure}[t]
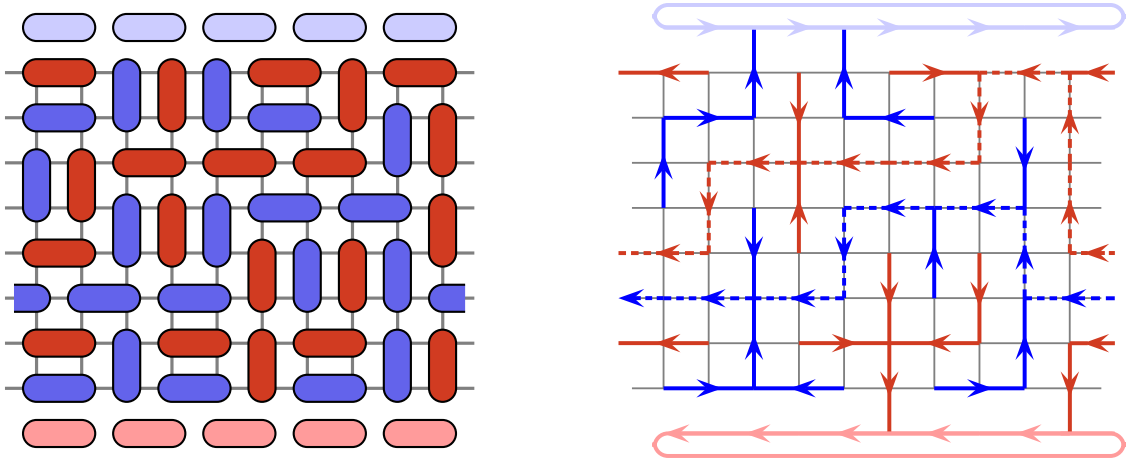

\psset{xunit=0.6cm}
\psset{yunit=0.6cm}
\psset{runit=0.6cm}
\def\hrhodimer{\psline[linearc=.30,linecolor=black,fillstyle=solid,fillcolor=myblue](0.5,-0.30)(-0.30,-0.30)(-0.30,0.30)(0.5,0.30)}
\def\hlhodimer{\psline[linearc=.30,linecolor=black,fillstyle=solid,fillcolor=myblue](-0.5,-0.30)(0.30,-0.30)(0.30,0.30)(-0.5,0.30)}
\def\hodimer{\pspolygon[linearc=.30,linecolor=black,fillstyle=solid,fillcolor=myblue](-0.30,-0.30)(-0.30,0.30)(1.30,0.30)(1.30,-0.30)}
\def\vodimer{\pspolygon[linearc=.30,linecolor=black,fillstyle=solid,fillcolor=myblue](-0.30,-0.30)(-0.30,1.30)(0.30,1.30)(0.30,-0.30)}
\def\hedimer{\pspolygon[linearc=.30,linecolor=black,fillstyle=solid,fillcolor=lightred](-0.30,-0.30)(-0.30,0.30)(1.30,0.30)(1.30,-0.30)}
\def\vedimer{\pspolygon[linearc=.30,linecolor=black,fillstyle=solid,fillcolor=lightred](-0.30,-0.30)(-0.30,1.30)(0.30,1.30)(0.30,-0.30)}
\pspicture(-3.5,-2)(10,9.5)
\multido{\nt=0+1}{8}{\psline[linewidth=1.2pt,linecolor=gray](-0.7,\nt)(9.7,\nt)}
\multido{\nt=0+1}{10}{\psline[linewidth=1.2pt,linecolor=gray](\nt,0)(\nt,7)}
\multido{\nt=0+2}{5}{\rput(\nt,-1){\pspolygon[linearc=.30,linecolor=black,fillstyle=solid,fillcolor=myred](-0.30,-0.30)(-0.30,0.30)(1.30,0.30)(1.30,-0.30)}}
\multido{\nt=0+2}{5}{\rput(\nt,8){\pspolygon[linearc=.30,linecolor=black,fillstyle=solid,fillcolor=lightblue](-0.30,-0.30)(-0.30,0.30)(1.30,0.30)(1.30,-0.30)}}
\rput(0,0){\hodimer}
\rput(2,0){\vodimer}
\rput(3,0){\hodimer}
\rput(5,0){\vedimer}
\rput(6,0){\hodimer}
\rput(8,0){\vodimer}
\rput(9,0){\vedimer}
\rput(0,1){\hedimer}
\rput(3,1){\hedimer}
\rput(6,1){\hedimer}
\rput(0,2){\hlhodimer}
\rput(1,2){\hodimer}
\rput(3,2){\hodimer}
\rput(5,2){\vedimer}
\rput(6,2){\vodimer}
\rput(7,2){\vedimer}
\rput(8,2){\vodimer}
\rput(9,2){\hrhodimer}
\rput(0,3){\hedimer}
\rput(2,3){\vodimer}
\rput(3,3){\vedimer}
\rput(4,3){\vodimer}
\rput(9,3){\vedimer}
\rput(0,4){\vodimer}
\rput(1,4){\vedimer}
\rput(5,4){\hodimer}
\rput(7,4){\hodimer}
\rput(2,5){\hedimer}
\rput(4,5){\hedimer}
\rput(6,5){\hedimer}
\rput(8,5){\vodimer}
\rput(9,5){\vedimer}
\rput(0,6){\hodimer}
\rput(2,6){\vodimer}
\rput(3,6){\vedimer}
\rput(4,6){\vodimer}
\rput(5,6){\hodimer}
\rput(7,6){\vedimer}
\rput(0,7){\hedimer}
\rput(5,7){\hedimer}
\rput(8,7){\hedimer}
\endpspicture
\hspace{2.2cm}
\pspicture(0,-2)(10,9.5)
\def\br{\psline[linewidth=1.5pt,linecolor=blue](1.,0)(2,0)
\psline[linewidth=1.5pt,linecolor=blue,arrowsize=4pt 2]{->}(0,0)(1.3,0)}
\def\bl{\psline[linewidth=1.5pt,linecolor=blue](-1,0)(-2,0)
\psline[linewidth=1.5pt,linecolor=blue,arrowsize=4pt 2]{->}(0,0)(-1.2,0)}
\def\bu{\psline[linewidth=1.5pt,linecolor=blue](0,1)(0,2)
\psline[linewidth=1.5pt,linecolor=blue,arrowsize=4pt 2]{->}(0,0)(0,1.2)}
\def\bd{\psline[linewidth=1.5pt,linecolor=blue](0,-1)(0,-2)
\psline[linewidth=1.5pt,linecolor=blue,arrowsize=4pt 2]{->}(0,0)(0,-1.2)}
\def\brd{\psline[linewidth=1.5pt,linestyle=dashed,dash=3pt 2pt,linecolor=blue](1.,0)(2,0)
\psline[linewidth=1.5pt,linestyle=dashed,dash=3pt 2pt,linecolor=blue,arrowsize=4pt 2]{->}(0,0)(1.3,0)}
\def\bld{\psline[linewidth=1.5pt,linestyle=dashed,dash=3pt 2pt,linecolor=blue](-1,0)(-2,0)
\psline[linewidth=1.5pt,linestyle=dashed,dash=3pt 2pt,linecolor=blue,arrowsize=4pt 2]{->}(0,0)(-1.2,0)}
\def\bud{\psline[linewidth=1.5pt,linestyle=dashed,dash=3pt 2pt,linecolor=blue](0,1)(0,2)
\psline[linewidth=1.5pt,linestyle=dashed,dash=3pt 2pt,linecolor=blue,arrowsize=4pt 2]{->}(0,0)(0,1.2)}
\def\bdd{\psline[linewidth=1.5pt,linestyle=dashed,dash=3pt 2pt,linecolor=blue](0,-1)(0,-2)
\psline[linewidth=1.5pt,linestyle=dashed,dash=3pt 2pt,linecolor=blue,arrowsize=4pt 2]{->}(0,0)(0,-1.2)}
\def\rr{\psline[linewidth=1.5pt,linecolor=lightred](1.,0)(2,0)
\psline[linewidth=1.5pt,linecolor=lightred,arrowsize=4pt 2]{->}(0,0)(1.3,0)}
\def\rl{\psline[linewidth=1.5pt,linecolor=lightred](-1,0)(-2,0)
\psline[linewidth=1.5pt,linecolor=lightred,arrowsize=4pt 2]{->}(0,0)(-1.2,0)}
\def\ru{\psline[linewidth=1.5pt,linecolor=lightred](0,1)(0,2)
\psline[linewidth=1.5pt,linecolor=lightred,arrowsize=4pt 2]{->}(0,0)(0,1.2)}
\def\rd{\psline[linewidth=1.5pt,linecolor=lightred](0,-1)(0,-2)
\psline[linewidth=1.5pt,linecolor=lightred,arrowsize=4pt 2]{->}(0,0)(0,-1.2)}
\def\rrd{\psline[linewidth=1.5pt,linestyle=dashed,dash=3pt 2pt,linecolor=lightred](1.,0)(2,0)
\psline[linewidth=1.5pt,linestyle=dashed,dash=3pt 2pt,linecolor=lightred,arrowsize=4pt 2]{->}(0,0)(1.3,0)}
\def\rld{\psline[linewidth=1.5pt,linestyle=dashed,dash=3pt 2pt,dash=3pt 2pt,linecolor=lightred](-1,0)(-2,0)
\psline[linewidth=1.5pt,linestyle=dashed,dash=3pt 2pt,dash=3pt 2pt,linecolor=lightred,arrowsize=4pt 2]{->}(0,0)(-1.2,0)}
\def\rud{\psline[linewidth=1.5pt,linestyle=dashed,dash=3pt 2pt,linecolor=lightred](0,1)(0,2)
\psline[linewidth=1.5pt,linestyle=dashed,dash=3pt 2pt,linecolor=lightred,arrowsize=4pt 2]{->}(0,0)(0,1.2)}
\def\rdd{\psline[linewidth=1.5pt,linestyle=dashed,dash=3pt 2pt,linecolor=lightred](0,-1)(0,-2)
\psline[linewidth=1.5pt,linestyle=dashed,dash=3pt 2pt,linecolor=lightred,arrowsize=4pt 2]{->}(0,0)(0,-1.2)}
\multido{\nt=0+1}{8}{\psline[linewidth=.7pt,linecolor=gray](-0.7,\nt)(9.7,\nt)}
\multido{\nt=0+1}{10}{\psline[linewidth=.7pt,linecolor=gray](\nt,0)(\nt,7)}
\rput(1,-1){\psline[linewidth=1.5pt,linecolor=myred,arrowsize=4pt 2]{->}(0,0)(-1,0)}
\rput(3,-1){\psline[linewidth=1.5pt,linecolor=myred](-1,0)(-2,0)
\psline[linewidth=1.5pt,linecolor=myred,arrowsize=4pt 2]{->}(0,0)(-1.2,0)}
\rput(5,-1){\psline[linewidth=1.5pt,linecolor=myred](-1,0)(-2,0)
\psline[linewidth=1.5pt,linecolor=myred,arrowsize=4pt 2]{->}(0,0)(-1.2,0)}
\rput(7,-1){\psline[linewidth=1.5pt,linecolor=myred](-1,0)(-2,0)
\psline[linewidth=1.5pt,linecolor=myred,arrowsize=4pt 2]{->}(0,0)(-1.2,0)}
\rput(9,-1){\psline[linewidth=1.5pt,linecolor=myred](-1,0)(-2,0)
\psline[linewidth=1.5pt,linecolor=myred,arrowsize=4pt 2]{->}(0,0)(-1.2,0)}
\rput(10,-1){\psline[linewidth=1.5pt,linecolor=myred,arrowsize=4pt 2](0,0)(-1.2,0)}
\rput(0,0){\br}
\rput(2,0){\bu}
\rput(4,0){\bl}
\rput(6,0){\br}
\rput(8,0){\bu}
\rput(1,1){\rl}
\rput(3,1){\rr}
\rput(5,1){\rd}
\rput(7,1){\rl}
\rput(9,1){\rd}
\psline[linewidth=1.5pt,linecolor=lightred,arrowsize=4pt 2]{->}(10,1)(9.3,1)
\psline[linewidth=1.5pt,linecolor=lightred,arrowsize=4pt 2](9.4,1)(9,1)
\rput(2,2){\bld}
\rput(4,2){\bld}
\rput(6,2){\bu}
\rput(8,2){\bud}
\rput(10,2){\bld}
\psline[linewidth=1.5pt,linestyle=dashed,dash=3pt 2pt,linecolor=blue,arrowsize=4pt 2]{->}(0,2)(-1,2)
\rput(1,3){\rld}
\rput(3,3){\ru}
\rput(5,3){\rd}
\rput(7,3){\rd}
\rput(9,3){\rud}
\psline[linewidth=1.5pt,linestyle=dashed,dash=3pt 2pt,linecolor=lightred,arrowsize=4pt 2]{->}(10,3)(9.3,3)
\psline[linewidth=1.5pt,linestyle=dashed,dash=3pt 2pt,linecolor=lightred,arrowsize=4pt 2](9.4,3)(9,3)
\rput(0,4){\bu}
\rput(2,4){\bd}
\rput(4,4){\bdd}
\rput(6,4){\bld}
\rput(8,4){\bld}
\rput(1,5){\rdd}
\rput(3,5){\rld}
\rput(5,5){\rld}
\rput(7,5){\rld}
\rput(9,5){\rud}
\rput(0,6){\br}
\rput(2,6){\bu}
\rput(4,6){\bu}
\rput(6,6){\bl}
\rput(8,6){\bd}
\rput(1,7){\rl}
\rput(3,7){\rd}
\rput(5,7){\rr}
\rput(7,7){\rdd}
\rput(9,7){\rld}
\psline[linewidth=1.5pt,linecolor=lightred,arrowsize=4pt 2]{->}(10,7)(9.3,7)
\psline[linewidth=1.5pt,linecolor=lightred,arrowsize=4pt 2](9.4,7)(9,7)
%
\rput(0,8){\psline[linewidth=1.5pt,linecolor=lightblue](1.,0)(2,0)
\psline[linewidth=1.5pt,linecolor=lightblue,arrowsize=4pt 2]{->}(0,0)(1.3,0)}
\rput(2,8){\psline[linewidth=1.5pt,linecolor=lightblue](1.,0)(2,0)
\psline[linewidth=1.5pt,linecolor=lightblue,arrowsize=4pt 2]{->}(0,0)(1.3,0)}
\rput(4,8){\psline[linewidth=1.5pt,linecolor=lightblue](1.,0)(2,0)
\psline[linewidth=1.5pt,linecolor=lightblue,arrowsize=4pt 2]{->}(0,0)(1.3,0)}
\rput(6,8){\psline[linewidth=1.5pt,linecolor=lightblue](1.,0)(2,0)
\psline[linewidth=1.5pt,linecolor=lightblue,arrowsize=4pt 2]{->}(0,0)(1.3,0)}
\rput(8,8){\psline[linewidth=1.5pt,linecolor=lightblue](1.,0)(2,0)
\psline[linewidth=1.5pt,linecolor=lightblue,arrowsize=4pt 2]{->}(0,0)(1.3,0)}
\pspolygon[linearc=.30,linecolor=myred,linewidth=1.5pt](-0.2,-1)(-0.2,-1.5)(10.2,-1.5)(10.2,-1)
\pspolygon[linearc=.30,linecolor=lightblue,linewidth=1.5pt](-0.2,8)(-0.2,8.5)(10.2,8.5)(10.2,8)
\endpspicture
\caption{The left figure shows a cylindrical dimer configuration on a $8 \times 10$ grid, with horizontal periodicity, where the blue (red) dimers are those which cover a site of ${\cal L}_{\rm odd}$ (${\cal L}_{\rm even}$). The right figure shows the corresponding arrow configurations as obtained from Temperley's correspondence. This particular configuration has a blue loop and a red loop in the bulk of the grid, wrapped around the cylinder and shown as dashed lines. The arrows flow towards the loops, and to roots, blue at the top, red at the bottom. Alternatively, one can border the dimer configuration with two rows of fixed horizontal dimers, shown in lightblue (top) and lightred (bottom) on the left figure. The two extra rows yield two extra loops, to which the arrows converge.}
\label{fig1}
\end{figure}

It is easy to see that every dimer covers exactly one site of either $\L_{\rm odd}$ or $\L_{\rm even}$, the other site being in the complement $\L \setminus (\L_{\rm odd} \cup \L_{\rm even})$. Accordingly one separates the dimers into two classes, odd or even, which we color in blue and red, as in Figure \ref{fig1}. 

Temperley's correspondence replaces each dimer by an arrow: the arrow starts from the site of $\L_{\rm odd}$ or $\L_{\rm even}$ which the dimer covers, is aligned along the dimer, and points to the next site of $\L_{\rm odd}$ or $\L_{\rm even}$, or towards the outside of the grid if the arrow goes out from a site which is a neighbour site of the boundary. In this case, we say that the arrow points to a root. Except in this last situation, the arrows connect two neighboring sites of $\L_{\rm odd}$ or of $\L_{\rm even}$. Finally, the arrows inherit the color from the dimers they are constructed from: they are colored blue if they originate from sites of $\L_{\rm odd}$, and red if they originate from sites of $\L_{\rm even}$. The set of blue arrows form a spanning web on $\L_{\rm odd}$, and the set of red arrows form a spanning web on $\L_{\rm even}$. For the rest, the complement $\L \setminus (\L_{\rm odd} \cup \L_{\rm even})$ plays no role and will be removed.

The boundary conditions are different on the two sublattices, as can be seen in Figure 1. Let us note that the lattice $\L_{\rm odd} \cup \L_{\rm even}$ consists of layers of sites of $\L_{\rm odd}$ alternating with layers of sites of $\L_{\rm even}$. 

If the number $M$ of rows is even, the bottom layer belongs to $\L_{\rm odd}$ and the top layer is in $\L_{\rm even}$. On $\L_{\rm odd}$, the blue arrows on the bottom row point to sites of $\L_{\rm odd}$, but those on the highest row of $\L_{\rm odd}$ (row number $M-1$) may also point vertically to a root. We will say that the boundary conditions on the bottom and top edges of $\L_{\rm odd}$ are respectively closed and open, see \cite{B95}. On $\L_{\rm even}$, the situation is reversed: the top edge (row $M$) is closed while the bottom edge of $\L_{\rm even}$ (row 2) is open.

If $M$ is odd, $\L_{\rm even}$ has one row less than $\L_{\rm odd}$. In this case, the two boundaries of $\L_{\rm odd}$ are closed, whereas the two boundaries of $\L_{\rm even}$ are open. 

If periodicity is imposed in one direction, say horizontally, we will always assume that the number $N$ of sites in that direction is even. Then $\L_{\rm odd}$ and $\L_{\rm even}$ inherit of the same periodicity in that direction. 

Thus the correspondence associates to each dimer configuration on a cylindrical grid two arrow configurations forming two spanning webs, a blue one and a red one. One can make a number of more or less elementary observations about the spanning webs that occur in this correspondence, all of which are well illustrated in Figure 1.

\begin{enumerate}
\item The arrows of the spanning webs, blue or red, flow to roots or to loops.

\item 
The blue and red spanning webs do not cross, but are intertwined; as $\L_{\rm even}$ is the dual lattice of $\L_{\rm odd}$, one can view the two spanning webs as dual.

\item A spanning web, either blue or red, never contains a contractible loop, because such a loop would encircle an odd number of sites of $\L$, which could therefore not be fully covered by dimers. If however periodicity is assumed in one direction, a spanning web may contain non-contractible loops wrapping around the periodic direction (like on Figure 1). Note that a non-contractible loop can wind around the cylinder only once, since it cannot intersect itself. A spanning web with no loop is a spanning forest. If the grid has no periodic direction, all spanning webs are spanning forests.

\item If a dimer configuration gives rise to an arrow configuration with a non-contractible loop, there is one other dimer configuration leading to identical arrows except that the orientation of the loop is opposite. This other dimer configuration is simply obtained by shifting the dimers forming the loop, by one lattice spacing along the loop. 

\item Suppose, like in Figure 1, that periodicity is assumed horizontally but not vertically. 

Then, if $M$ is even, the blue spanning web contains the same number of loops as the red spanning web; moreover blue and red loops alternate from bottom to top, starting with a blue one and ending with a red one. Indeed if the blue spanning web contains a loop, the red arrows located above the blue loop cannot flow to roots because the top boundary of $\L_{\rm even}$ is closed. Therefore they must be a red loop to which they can flow.

If $M$ is odd, the blue spanning web has one more loop than the red spanning web. The loops of either color still alternate, starting and ending with a blue one. The reason for this is that the two boundaries of $\L_{\rm odd}$ are closed, so that the blue arrows have nowhere to flow except to loops.

\item An important observation for what follows is that the dimers of one color unambiguously determines the positions of the dimers of the other color, up to the orientation of the loops. As therefore follows from the previous remarks, to a fixed configuration of blue arrows containing $L$ loops, there are $2^L$ or $2^{L-1}$ (depending on the parity of $M$) possible arrangements of the red arrows, equivalently, of the dimers touching sites of $\L_{\rm even}$.

\item The correspondence between dimer configurations and the pair (blue and red) of oriented spanning webs of the sort described above is one-to-one, but is in general many-to-one if one keeps the spanning web of only one color, because of the orientation of the loops. 

\end{enumerate}

The correspondence recalled above allows to trade the statistical analysis of certain webs on a cylindrical grid, taken as the $\L_{\rm odd}$ of $\L_{\rm even}$ of some larger cylindrical grid $\L$, to a related analysis of dimer configurations on $\L$. By choosing an appropriate measure on the dimer configurations, one can disentangle the different sorts of spanning webs that occur. As the multiplicity introduced by the orientations of the loops is known, one may choose to consider oriented spanning webs or else to ignore the orientation. 

The dimer setting is convenient to work with, especially in view of the fact that it allows for a transfer matrix treatment. In the rest of this article, we will show how the different classes of spanning webs can be separated, and also how the loops can be removed, therefore leading to a situation where spanning forests only remain. As the recurrent configurations of the sandpile model are equivalent to spanning forests \cite{dharmaj}, this will provide a transfer matrix for the sandpile model defined on a cylinder (but not on a torus).


\section{Counting dimers}
\label{sec3}

The construction of a transfer matrix for dimer configurations has been given long ago by Lieb \cite{lieb}. A recent account of it can also be found in \cite{rasru}.

As explained above, we consider all possible dimer configurations on a rectangular grid in $\Z^2$ with $M$ rows and $N$ columns. For the moment, periodicity may be assumed horizontally or vertically, or both. In this section we restrict to the probability measure that assigns a dimer configuration the statistical weight $\alpha^h \beta^v$ discussed in the previous section. We will see in the next section how to modify it in order to keep track of the loops in the spanning web picture. The partition function sums these weights for all configurations, but since $h+v=MN/2$, it essentially depends only on the ratio $\alpha/\beta$. Equivalently, setting $\beta=1$ without loss of generality, we can write
\be
{\cal Z}_{M,N} = \sum_{\rm configs} \; \a^h.
\ee

\begin{figure}[b]
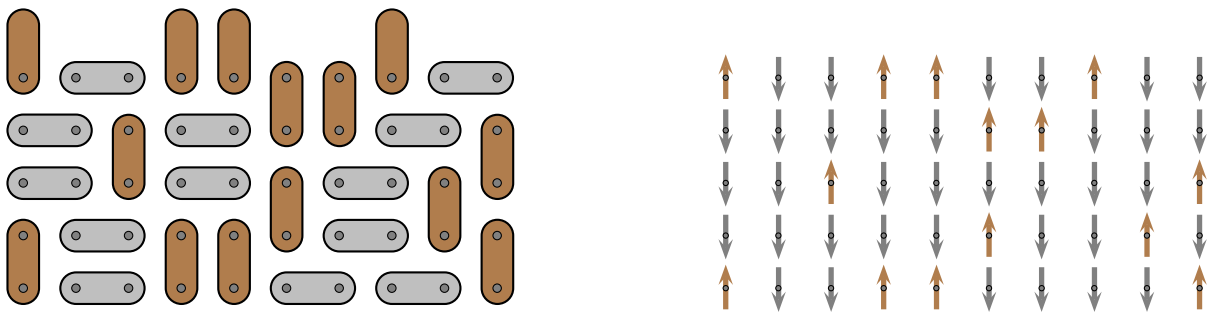

\psset{xunit=0.7cm}
\psset{yunit=0.7cm}
\psset{runit=0.7cm}
\def\hdimer{\pspolygon[linearc=.30,linecolor=black,fillstyle=solid,fillcolor=lightgray](-0.30,-0.30)(-0.30,0.30)(1.30,0.30)(1.30,-0.30)}
\def\vdimer{\pspolygon[linearc=.30,linecolor=black,fillstyle=solid,fillcolor=mybrown](-0.30,-0.30)(-0.30,1.30)(0.30,1.30)(0.30,-0.30)}
\pspicture(-1.5,-1)(10,6.5)
\rput(0,0){\vdimer}
\rput(1,0){\hdimer}
\rput(3,0){\vdimer}
\rput(4,0){\vdimer}
\rput(5,0){\hdimer}
\rput(7,0){\hdimer}
\rput(9,0){\vdimer}
\rput(1,1){\hdimer}
\rput(5,1){\vdimer}
\rput(6,1){\hdimer}
\rput(8,1){\vdimer}
\rput(0,2){\hdimer}
\rput(2,2){\vdimer}
\rput(3,2){\hdimer}
\rput(6,2){\hdimer}
\rput(9,2){\vdimer}
\rput(0,3){\hdimer}
\rput(3,3){\hdimer}
\rput(5,3){\vdimer}
\rput(6,3){\vdimer}
\rput(7,3){\hdimer}
\rput(0,4){\vdimer}
\rput(1,4){\hdimer}
\rput(3,4){\vdimer}
\rput(4,4){\vdimer}
\rput(7,4){\vdimer}
\rput(8,4){\hdimer}
\multido{\nt=0+1}{10}{\pscircle[linewidth=0.4pt,fillstyle=solid,fillcolor=gray](\nt,0){0.09}}
\multido{\nt=0+1}{10}{\pscircle[linewidth=0.4pt,fillstyle=solid,fillcolor=gray](\nt,1){0.09}}
\multido{\nt=0+1}{10}{\pscircle[linewidth=0.4pt,fillstyle=solid,fillcolor=gray](\nt,2){0.09}}
\multido{\nt=0+1}{10}{\pscircle[linewidth=0.4pt,fillstyle=solid,fillcolor=gray](\nt,3){0.09}}
\multido{\nt=0+1}{10}{\pscircle[linewidth=0.4pt,fillstyle=solid,fillcolor=gray](\nt,4){0.09}}
\endpspicture
\hspace{2.2cm}
\pspicture(0,-1)(10,6.5)
\rput(0,-0.5){
\rput(0,0){\aup}
\rput(1,0){\ado}
\rput(2,0){\ado}
\rput(3,0){\aup}
\rput(4,0){\aup}
\rput(5,0){\ado}
\rput(6,0){\ado}
\rput(7,0){\ado}
\rput(8,0){\ado}
\rput(9,0){\aup}
\rput(0,1){\ado}
\rput(1,1){\ado}
\rput(2,1){\ado}
\rput(3,1){\ado}
\rput(4,1){\ado}
\rput(5,1){\aup}
\rput(6,1){\ado}
\rput(7,1){\ado}
\rput(8,1){\aup}
\rput(9,1){\ado}
\rput(0,2){\ado}
\rput(1,2){\ado}
\rput(2,2){\aup}
\rput(3,2){\ado}
\rput(4,2){\ado}
\rput(5,2){\ado}
\rput(6,2){\ado}
\rput(7,2){\ado}
\rput(8,2){\ado}
\rput(9,2){\aup}
\rput(0,3){\ado}
\rput(1,3){\ado}
\rput(2,3){\ado}
\rput(3,3){\ado}
\rput(4,3){\ado}
\rput(5,3){\aup}
\rput(6,3){\aup}
\rput(7,3){\ado}
\rput(8,3){\ado}
\rput(9,3){\ado}
\rput(0,4){\aup}
\rput(1,4){\ado}
\rput(2,4){\ado}
\rput(3,4){\aup}
\rput(4,4){\aup}
\rput(5,4){\ado}
\rput(6,4){\ado}
\rput(7,4){\aup}
\rput(8,4){\ado}
\rput(9,4){\ado}}
\multido{\nt=0+1}{10}{\pscircle[linewidth=0.4pt,fillstyle=solid,fillcolor=gray](\nt,0){0.06}}
\multido{\nt=0+1}{10}{\pscircle[linewidth=0.4pt,fillstyle=solid,fillcolor=gray](\nt,1){0.06}}
\multido{\nt=0+1}{10}{\pscircle[linewidth=0.4pt,fillstyle=solid,fillcolor=gray](\nt,2){0.06}}
\multido{\nt=0+1}{10}{\pscircle[linewidth=0.4pt,fillstyle=solid,fillcolor=gray](\nt,3){0.06}}
\multido{\nt=0+1}{10}{\pscircle[linewidth=0.4pt,fillstyle=solid,fillcolor=gray](\nt,4){0.06}}
\endpspicture
\caption{Lieb's association of spins to dimers. The color code on the left figure simply refers to whether a dimer is horizontal or vertical. The right figure shows the resulting spin configurations.}
\label{fig2}
\end{figure}

The idea underlying Lieb's construction is to code the presence of a horizontal or vertical dimer in terms of up and down spins, in such a way that a whole configuration of dimers can be traded for a configuration of spins. The spins are attached to the sites, see the figures below. An up spin at site $i$ means that there is a vertical dimer covering $i$ and its top neighbour; a down spin indicates the absence of such a vertical dimer. In the latter case, it either means, if the down spin is right above an up spin, that site $i$ pairs up with its neighbour below to form a vertical dimer; otherwise, it means the down spin at $i$ is next to another down spin, at the left or at the right neighbouring site, to form a horizontal dimer. A row of $N$ sites can have $2^N$ different spin configurations. The correspondence is pictured in Fig. \ref{fig2} for a dimer configuration covering five successive rows. 

The transfer matrix iteratively builds up the dimer configuration row by row, starting from a first row, at the bottom of the grid. It produces a new row of spins from the previous one (the one below it) by making appropriate flips. As is customary, we can think of the up and down spins as the canonical base elements $|\!\!\uparrow\ra = \scriptsize (\matrix{1 \cr 0})$ and $|\!\!\downarrow\ra = \scriptsize (\matrix{0 \cr 1})$ of a vector space $\C^2$, so that a row configuration of spins is an element of $(\C^2)^{\otimes N}$. In this case the different actions on spins are realized by Pauli matrices $\s_i = {\mathbb I} \otimes \ldots \otimes \s \otimes \ldots \otimes {\mathbb I}$, where $\s$, in position $i$, is one the following matrices 
\be
\s^x = \pmatrix{0 & 1 \cr 1 & 0}, \qquad \s^- = \pmatrix{0 & 0 \cr 1 & 0}, \qquad\s^+ = \pmatrix{0 & 1 \cr 0 & 0}.
\ee
The matrices on different sites commute, $\s_i\s_j=\s_j\s_i$ for all $i \neq j$, while those at the same site satisfy $(\s_i^-)^2 = (\s_i^+)^2 = (\s_i^x)^2-{\mathbb I} = 0$ and $\s_i^x \, \s_i^+ \, \s_i^x = \s_i^-$.

From the way the association dimers-spins is defined, a down spin can be followed on the next row by a down or an up spin, but an up spin can only be followed by a down spin. It follows that the transfer matrix reads \cite{lieb}
\be
T = \prod_i\: [\bi + \a\, \s_i^-\s_{i+1}^-] \: \prod_{i=1}^N \: \s_i^x = \exp{(\sum_i \, \a\,\s_i^-\s_{i+1}^-)} \, \prod_{i=1}^N \: \s_i^x.
\label{tm}
\ee

It is not difficult to see that, when acting on an incoming horizontal sequence of spins, this matrix does exactly what it should. First the factor $V_1 = \prod_i \, \s_i^x$ flips all the incoming spins, and makes sure that an up spin goes out as a down spin. This is followed by the action of factors $\bi + \a \s_i^-\s_{i+1}^-$. Each such factor either does nothing ($\bi$) or else turns two up spins at $i$ and $i+1$ into down spins to form a horizontal dimer, and at the same time assigns the correct weight $\a$. The product over $i$ produces this alternative for every pair of neighbouring sites in the row.

Thus the matrix $T$ outputs all possible spin configurations from an incoming row $|{\rm in}\ra$ while respecting the constraints: it correctly handles the incoming vertical dimers, and creates horizontal dimers wherever possible. Likewise, the power $T^m$ successively constructs all possible arrays of $n$ rows built on top of $|{\rm in}\rangle$ which satisfy the rules prescribed by the model. The top row in this process, $|{\rm out}\rangle = T^m |{\rm in}\rangle$, gives all possible outgoing row configurations, $m$ layers higher than the incoming row $|{\rm in}\rangle$. 

Two different boundary conditions can be considered on the vertical edges, straight and periodic. Straight means that the grid is bordered, on the left and on the right, by straight edges which form barriers holding the dimers (horizontally the geometry is that of a strip). In this case the summation over $i$ in the exponential in (\ref{tm}) runs from 1 to $N-1$, whereas it includes $N$ in the periodic case, as well as the periodicity condition $\s_{N+1} = \s_1$. The only difference in the periodic case is, therefore, an extra term $\a\,\s_1^-\s_N^-$ in the exponential. 

The same two boundary conditions can also be considered on the bottom and top boundaries, but the way they are implemented is different. If straight boundary conditions are assumed, the incoming bottom row $|{\rm in}\ra$ as well as the outgoing top row $|{\rm out}\ra$ are constrained. A spin up in $|{\rm in}\ra$ means that a vertical dimer is located at that position, pointing upwards; a down spin indicates the absence of a vertical dimer, and therefore means that this site is covered by a horizontal dimer which also covers a neighbouring site, itself with a down spin. Therefore down spins must come in pairs. On the other hand, a spin up in $|{\rm out}\ra$ is forbidden, as it would mean that a vertical dimer is going upwards, and would therefore stick out from the top row. Thus the outgoing state must be the all down spin state, $|{\rm out}\ra = |\!\downarrow\downarrow \ldots \downarrow\ra$.

We obtain, for straight boundary conditions on the bottom and top rows (the choice of $T$, as discussed above, determines the horizontal boundary conditions), 
\be
{\cal Z}_{M,N} = \sum_{|{\rm in\ra}} \; \la\downarrow\downarrow \ldots \downarrow\!|\:T^{M-1}\:|{\rm in}\ra,
\ee
where the states $|{\rm in}\ra$ are constrained as explained above, and in addition, must be weighted by the appropriate power of $\alpha$ according to the number of pairs of (adjacent) spins down they contain.

One may simplify this expression by observing that the allowed incoming states are exactly all the possible states output by the transfer matrix when it acts on the all down spin configuration, and that they automaticaly get the correct weight. We then obtain the nicer formula,
\be
{\cal Z}_{M,N} = \la\downarrow\downarrow \ldots \downarrow\!|\:T^M\:|\!\downarrow\downarrow \ldots \downarrow\ra. \qquad\quad \hbox{(straight bottom and top boundaries)}
\label{str}
\ee

If vertical periodicity is assumed, the row number $M+1$, produced by the action of $T^M$, is to be identified with the initial row $|{\rm in}\ra$, so that the partition function takes the form
\be
{\cal Z}^{}_{M,N} = \sum_{|{\rm in\ra}} \; \langle {\rm in}|\, T^M \,|{\rm in}\rangle.
\ee
In this case, there is no restriction on the initial states $|{\rm in}\ra$. An isolated down spin in $|{\rm in}\ra$ is allowed and corresponds to a vertical dimer coming from the previous row, in this case from a site of the top ($M$-th) row. Therefore we get the familiar formula
\be
{\cal Z}^{}_{M,N} = {\rm Tr}\: T^M. \qquad\qquad \hbox{(periodic condition vertically)}
\ee
Let us note that partial traces can also be considered. As shown in \cite{rasru}, the full configuration space of spins on a row, of dimension $2^N$, can be divided in disjoint sectors, each sector being left invariant by the action of $T$. This splitting stems from the existence of an operator $\cal V$ which anticommutes with the transfer matrix and is diagonal in Lieb's spin basis,
\be
{\cal V} = {1 \over 2} \: \sum_{i=1}^N \; (-1)^i \: \sigma_i^z, \qquad \sigma^z = \pmatrix{1 & 0 \cr
0 & -1}.
\label{nu}
\ee
Its eigenvalues, called variation indices in \cite{rasru}, vary between $-{N \over 2}$ and $N \over 2$ by integer steps and label the different sectors. The partial traces restricted to sectors lead to refined partition functions. 


\section{Handling the loops}
\label{sec4}

Our primary purpose is to study the spanning webs arising from Temperley's correspondence with dimers on a cylindrical grid, and to separate them according to the number of loops they contain. As there are no loops if no direction is periodic, we assume horizontal, but not vertical, periodicity, and take $N$ to be even. 

\begin{figure}[t]
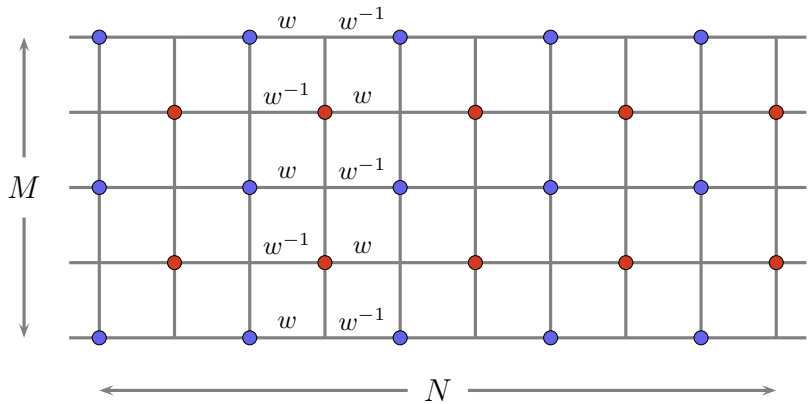

\psset{xunit=1cm}
\psset{yunit=1cm}
\psset{runit=1cm}
\begin{center}
\pspicture(-1.5,-0.5)(10,5)
\multido{\nt=0+1}{5}{\psline[linewidth=1.2pt,linecolor=gray](-0.4,\nt)(9.4,\nt)}
\multido{\nt=0+1}{10}{\psline[linewidth=1.2pt,linecolor=gray](\nt,0)(\nt,4)}
\multido{\ny=0.2+2.0}{3}{\multido{\nx=2.5+2.0}{1}{\rput(\nx,\ny){\footnotesize $w$}}}
\multido{\ny=0.25+2.00}{3}{\multido{\nx=3.5+2.0}{1}{\rput(\nx,\ny){\footnotesize $w^{-1}$}}}
\multido{\ny=1.25+2.00}{2}{\multido{\nx=2.5+2.0}{1}{\rput(\nx,\ny){\footnotesize $w^{-1}$}}}
\multido{\ny=1.2+2.0}{2}{\multido{\nx=3.5+2.0}{1}{\rput(\nx,\ny){\footnotesize $w$}}}
\multido{\ny=0+2}{3}{\multido{\nx=0+2}{5}{\pscircle[linewidth=0.4pt,fillstyle=solid,fillcolor=myblue](\nx,\ny){0.1}}}
\multido{\ny=1+2}{2}{\multido{\nx=1+2}{5}{\pscircle[linewidth=0.4pt,fillstyle=solid,fillcolor=lightred](\nx,\ny){0.1}}}
\rput(4.5,-0.7){$N$}
\psline[linewidth=1.2pt,linecolor=gray]{->}(4.1,-0.7)(0,-0.7)
\psline[linewidth=1.2pt,linecolor=gray]{->}(4.9,-0.7)(9,-0.7)
\rput(-1,2){$M$}
\psline[linewidth=1.2pt,linecolor=gray]{->}(-1,1.6)(-1,0)
\psline[linewidth=1.2pt,linecolor=gray]{->}(-1,2.4)(-1,4)
\endpspicture
\end{center}
\caption{The figure shows the weights associated to horizontal dimers used in the text to disentangle the configurations with different number of loops. As before, the sites of $\L_{\rm odd}$ resp. $\L_{\rm even}$ are colored in blue resp. red.}
\label{fig3}
\end{figure}

We do not want to make a distinction between horizontal and vertical dimers, as in the previous section, so we take $\alpha=1$. However we want to be able to distinguish the dimer configurations which give rise to spanning webs with different numbers of loops. A simple way to achieve this is pictured in Figure \ref{fig3}: we assign alternating weights $w$ and $w^{-1}$ to the horizontal bonds contained in two columns in the way indicated. All other horizontal bonds and all vertical bonds keep a weight equal to 1.

Thus a dimer has generically a weight equal to 1, but will get a weight $w$ or $w^{-1}$ if it covers one of the marked horizontal bonds. As a consequence, a sequence of dimers forming a loop oriented from left to right,  either blue or red, will get a weight equal to $w$. A loop with the opposite orientation, from right to left, will get a weight $w^{-1}$. More generally, a dimer configuration leading to a spanning web with $n_+$ loops oriented left-right and $n_-$ loops oriented right-left has a weight equal to $w^{n_+ - n_-}$. Let us write the dimer partition function in terms of the loop content of the associated spanning webs.


If $M$ is even, the number of blue loops is equal to the number of red loops. According to the above discussion, a dimer configuration with a total of $2L$ loops ($L$ blue and $L$ red) contributes a factor $w^{n_+-n_-}$, with $n_+ + n_- = 2L$. The partition function is thus
\be
{\cal Z}^{}_{M,N}(w) = \sum_{L=0}^{M/2} \; \sum_{n_+=0}^{2L} \; \#\Big[{\hbox{\small configurations with $2L$ loops,} \atop {\hbox{\small $n_+$ loops oriented left-right}}}\Big] \: w^{n_+ - n_-}.
\ee
The case $n_+=2L$ corresponds to the dimer configurations with all loops oriented left-right. For each such configuration, one can reverse the orientation of $n_-$ loops by moving the dimers along loops, as explained in Section 2. As there are ${2L \choose n_-} = {2L \choose n_+}$ ways to choose $n_-$ loops out of $2L$, we obtain
\be
\#\Big[{\hbox{\small configurations with $2L$ loops,} \atop {\hbox{\small $n_+$ loops oriented left-right}}}\Big] = {2L \choose n_+} \; \#\Big[{\hbox{\small configurations with $2L$ loops,} \atop {\hbox{\small all loops oriented left-right}}}\Big], 
\ee
and the partition function becomes
\be
{\cal Z}^{}_{M,N}(w) = \sum_{L=0}^{M/2} \; \#\Big[{\hbox{\small configurations with $2L$ loops,} \atop {\hbox{\small all loops oriented left-right}}}\Big] \: (w + w^{-1})^{2L}.
\ee
As we do not want to keep track of the orientation of the loops but only their number, we may write
\be
{\cal Z}^{}_{M,N}(w) = \sum_{L=0}^{M/2} \; \#\Big[{\hbox{\small configurations with} \atop {\hbox{\small $2L$ oriented loops}}}\Big] \; \Big({w + w^{-1} \over 2} \Big)^{2L}.
\qquad \hbox{($M$ even)}
\label{loop1}
\ee
It is to be emphasized that even though we do not keep track of the orientations, the loops are oriented. This means that two spanning webs differing by the orientation of some of their loops correspond to distinct dimer configurations and therefore are counted as being different. A natural alternative is to consider non-oriented loops, in which spanning webs are identified, and counted as one, if they only differ by the orientation of some of their loops. This means that we keep only one representative among $2^{2L}$ oriented spanning webs, leading to
\be
{\cal Z}^{}_{M,N}(w) = \sum_{L=0}^{M/2} \; \#\Big[{\hbox{\small configurations with} \atop {\hbox{\small $2L$ non-oriented loops}}}\Big] \; (w + w^{-1})^{2L}.
\qquad \hbox{($M$ even)}
\label{loop2}
\ee

As illustration, for a $4 \times 4$ cylinder with horizontal periodicity, one finds 
\be
{\cal Z}_{4,4}(w) = 29 + 19 (w + w^{-1})^2 + (w + w^{-1})^4,
\ee
and thus 29 spanning trees, 19 configurations with two loops oriented left-right (or as many non-oriented loops), and 1 configuration with four loops oriented left-right.

It is also noted that the value $w = {\rm i}$ satisfying $w + w^{-1} = 0$ automatically removes all the configurations with loops: the corresponding partition function counts the configurations which are spanning forests, and assigns a zero weight to the configurations containing loops. 

If $M$ is odd, there is always one more blue loop than red loops, so the total number of loops is odd. In this case the partition function reads
\bea
{\cal Z}^{}_{M,N}(w) &=& \sum_{L=0}^{(M-1)/2} \: \#\Big[{\hbox{\small configurations with $2L+1$} \atop \hbox{\small oriented loops, $L+1$ blue}}\Big] \; \Big({w + w^{-1} \over 2} \Big)^{2L+1}
\quad \hbox{($M$ odd)}\\
&=& \sum_{L=0}^{(M-1)/2} \: \#\Big[{\hbox{\small configurations with $2L+1$} \atop \hbox{\small non-oriented loops, $L+1$ blue}}\Big] \; (w + w^{-1})^{2L+1}.
\eea
It implies that the number of spanning trees on $\L_{\rm odd}$ is equal to 0, as expected since the boundary conditions on $\L_{\rm odd}$ are both closed, and that the number of spanning forests on $\L_{\rm even}$ (boundary conditions on $\L_{\rm even}$ are both open) is the coefficient of $(w + w^{-1})$. For instance, one finds 
\be
Z_{5,6}(w) = 1728\,(w + w^{-1}) + 124\, (w + w^{-1})^3 + (w + w^{-1})^5,
\ee
implying that there are $1728$ spanning forests on $\L_{\rm even}$, a $2 \times 3$ cylindrical grid with two open edges.


\section{Loops on a cylinder}
\label{sec5}

We have shown in the previous sections that the spanning webs on a cylinder $\cal C$ with no other loops than those wrapping around the cylinder can be traded for dimer configurations on an extended cylindrical lattice, weighted appropriately so that the spanning webs with different number of loops can be separated. Depending on the boundary conditions chosen on the two edges, open or closed, the cylinder should be identified as the even or odd lattice of the extended one. 

If the extended dimer lattice $\L$ is $2M \times 2N$ (even height, even perimeter), the odd and even sublattices, both $M \times N$, are subjected to the same boundary conditions : one edge is open, the other one is closed. In this case, the two sublattices play a symmetrical role, and we may take $\cal C$ to be either of them, say ${\cal C} = \L_{\rm even}$. From the previous section, we have
\be
{\cal Z}_{2M,2N}(w) = \sum_{L=0}^{M} \; \#\Big[{\hbox{\small configurations with} \atop {\hbox{\small $2L$ oriented loops}}}\Big] \: \Big({w + w^{-1} \over 2} \Big)^{2L}.
\ee

Each dimer configuration on $\L$ with $2L$ loops gives rise to a spanning web on $\cal C$ with $L$ loops, and another one on $\L_{\rm odd}$, also with $L$ loops. Because of the possible orientations of the $L$ loops on $\L_{\rm odd}$, we find that for each spanning web on $\cal C$ with $L$ loops, there are $2^L$ configurations on $\L$,
\bea
Z_{\rm op,cl}^{(L \,\rm loops)}(M,N) &=& \#[\hbox{oriented spanning webs on $\cal C$ with $L$ loops}]\nonumber\\
\egal 2^{-L} \cdot \#[\hbox{dimer configurations on $\L$ with $2L$ loops}].
\eea
We obtain the generating function for the $Z_{\rm op,cl}^{(L \,\rm loops)}$ as
\be
{\cal Z}_{2M,2N}(w) = \sum_{L=0}^M \; Z_{\rm op,cl}^{(L \,\rm loops)}(M,N) \: \Big({w + w^{-1} \over \sqrt{2}} \Big)^{2L}.
\ee

If the dimer lattice is $(2M+1) \times 2N$ (odd height, even perimeter), then $\L_{\rm odd}$ is $(M+1) \times N$ and has two closed edges, while $\L_{\rm even}$ is $M \times N$ and has two open edges. A dimer configuration on $\L$ with $2L+1$ loops gives rise to a spanning web on $\L_{\rm odd}$ with $L+1$ loops, and another one on $\L_{\rm even}$, with $L$ loops.  Focusing on either sublattice, and owing to the possible orientations of the loops on the other sublattice, we find 
\bea
&& \hspace{-7mm} Z_{\rm op,op}^{(L \,\rm loops)}(M,N) = 2^{-(L+1)} \cdot \#[\hbox{dimer configurations on $\L$ with $2L+1$ loops}],\\
\noalign{\medskip}
&& \hspace{-7mm} Z_{\rm cl,cl}^{(L+1 \,\rm loops)}(M+1,N) = 2^{-L} \cdot \#[\hbox{dimer configurations on $\L$ with $2L+1$ loops}],
\eea
implying $Z_{\rm cl,cl}^{(L+1 \,\rm loops)} = 2\, Z_{\rm op,op}^{(L \,\rm loops)}$ in the scaling limit. We obtain their generating function as
\bea
&& \hspace{-7mm} {\cal Z}_{2M+1,2N}(w) = \sum_{L=0}^M \: \#\Big[{\hbox{\small configurations with $2L+1$} \atop \hbox{\small oriented loops, $L+1$ blue}}\Big] \; \Big({w + w^{-1} \over 2} \Big)^{2L+1}\\
\noalign{\medskip}
&& \hspace{16mm}  = \sum_{L=0}^M \; Z_{\rm op,op}^{(L \,\rm loops)} \: {(w + w^{-1})^{2L+1} \over 2^L} = \sum_{L=0}^M \; Z_{\rm cl,cl}^{(L+1 \,\rm loops)} \: {(w + w^{-1})^{2L+1} \over 2^{L+1}}.
\eea

Thus all partial partition functions $Z^{(L \,\rm loops)}$ pertaining to the spanning webs containing a fixed number of loops and given boundary conditions can be computed from a dimer partition function, provided the dimer configurations are appropriately weighted, in shown in Figure \ref{fig3}. This can be done with the help of the transfer matrix as follows. 

\subsection{Weighted dimer configurations}

\begin{figure}[t]
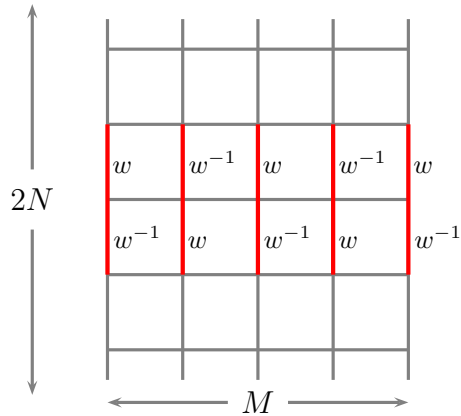

\psset{xunit=1cm}
\psset{yunit=1cm}
\psset{runit=1cm}
\begin{center}
\pspicture(-1.5,-0.5)(5,5)
\multido{\nt=0+1}{5}{\psline[linewidth=1.2pt,linecolor=gray](0,\nt)(4,\nt)}
\multido{\nt=0+1}{5}{\psline[linewidth=1.2pt,linecolor=gray](\nt,-0.4)(\nt,4.4)}
\multido{\nt=0+1}{5}{\psline[linewidth=1.6pt,linecolor=red](\nt,1)(\nt,3)}
\multido{\nx=0.4+2.0}{3}{\rput(\nx,1.5){\footnotesize $w^{-1}$}}
\multido{\nx=0.2+2.0}{3}{\rput(\nx,2.43){\footnotesize $w$}}
\multido{\nx=1.4+2.0}{2}{\rput(\nx,2.5){\footnotesize $w^{-1}$}}
\multido{\nx=1.2+2.0}{2}{\rput(\nx,1.43){\footnotesize $w$}}
\rput(2,-0.7){$M$}
\psline[linewidth=1.2pt,linecolor=gray]{->}(1.6,-0.7)(0,-0.7)
\psline[linewidth=1.2pt,linecolor=gray]{->}(2.4,-0.7)(4,-0.7)
\rput(-1,2){$2N$}
\psline[linewidth=1.2pt,linecolor=gray]{->}(-1,1.6)(-1,-0.6)
\psline[linewidth=1.2pt,linecolor=gray]{->}(-1,2.4)(-1,4.6)
\endpspicture
\end{center}
\caption{Weights associated to vertical bonds, as obtained from Figure \ref{fig3} rotated by 90 degrees.}
\label{fig4}
\end{figure}

The most convenient way to compute the required dimer partition functions with the transfer matrix is to choose the periodic direction to be the direction of transfer. The marked bonds, weighted with $w$ or $w^{-1}$, then appeared as vertical bonds, as pictured in Figure \ref{fig4}. Away from these, we use the transfer matrix introduced in Section \ref{sec3} (the left and right edges are straight)
\be
T = \prod_{i=1}^{M-1}\: [\bi + \s_i^-\s_{i+1}^-] \: \prod_{i=1}^M \: \s_i^x = \exp{(\sum_{i=1}^{M-1} \,\s_i^-\s_{i+1}^-)} \, \prod_{i=1}^M \: \s_i^x,
\label{t}
\ee
where $M$ can be even or odd.

Let us denote by $T(w)$ the modified transfer matrix that has to be used when the first row of weighted vertical bonds is encountered. Then, the matrix to be used immediately after $T(w)$ is $T(w^{-1})$ so that the dimer partition function reads
\be
{\cal Z}_{M,2N} = {\rm Tr}\, \Big[T^{2N-2} \, T(w^{-1}) \, T(w)\Big].
\ee

The modified matrix $T(w)$ should assign an outgoing up spin a factor $w$ or $w^{-1}$, depending on its position. The operator $w^{{1 \over 2}(\bi + \s_i^z)}$ precisely produces a factor $w$ if the outgoing spin at $i$ is up, and a factor 1 if it is down. Therefore we obtain
\be
T(w) = \Big\{\prod_{i=1}^M \, w^{(-1)^i (\bi + \s_i^z)/2} \Big\} \; T = w^{p(M)} \, w^{\cal V} \, T, \qquad p(M) = {1-(-1)^M \over 4},
\ee
where $\cal V$ is the variation operator introduced before in (\ref{nu}). Recalling that $\cal V$ anticommutes with $T$ \cite{rasru}, the partition function takes the form
\be
{\cal Z}_{M,2N} = {\rm Tr}\, \Big[T^{2N-2} \, w^{-\cal V}\,T \, w^{\cal V}\,T\Big] = {\rm Tr}\, \Big[T^{2N} w^{2\cal V}\Big] = {\rm Tr}\, \Big[T^{2N} w^{-2\cal V}\Big].
\ee
Since the operator $\cal V$, like $T$, acts on rows of spins of length $M$, its eigenvalues vary between $-{M \over 2}$ and $M \over 2$. Consequently, the trace in the previous equation is a finite series in $w$ with powers ranging between $-M$ and $M$, and includes only powers of the same parity as $M$.
These traces, for $M$ even or odd, have been computed in \cite{rasru}. In what follows, we set $w = {\rm e}^{{\rm i}\pi z}$.

\subsection{Mixed open/closed boundary conditions}

As discussed above, this case corresponds to an extended cylindrical lattice with an even height, which we denote by $2M$. The result is \cite{rasru}
\be
{\cal Z}_{2M,2N}(z) = {\theta_3(e^{2{\rm i}\pi z}|q) \over \eta(q)}\,, \qquad q =  \exp{(-\pi N/M)}.
\ee
For $z=0$ (dimers), it reduces to ${\theta_3 \over \eta}(q)$, in agreement with an old known result \cite{MW73,LuWu99}. For $z={1 \over 2}$, namely $w = \rm i$ corresponding to spanning forests or equivalently to the recurrent configurations of the sandpile model, one finds ${\theta_4 \over \eta}(q)$, reproducing an earlier result \cite{pr2002}. 

In the scaling limit, we may extract the partition functions $Z_{\rm op,cl}^{(L \,\rm loops)}$ for spanning webs involving a fixed number of loops by using the $q$-series expansion of $\theta_3$, 
\be
\hspace{-5mm}
{\cal Z}_{2M,2N}(z) = \sum_{L=0}^\infty \; Z_{\rm op,cl}^{(L \,\rm loops)}(q) \: (\sqrt{2} \, \cos{\pi z})^{2L} = {1 \over \eta(q)} \Big\{ 1 + 2 \sum_{n=1}^\infty \; \cos{2\pi nz} \: q^{n^2/2}\Big\}.
\ee
Using the Tchebychev polynomials $T_{2n}(\cos{\pi z}) = \cos{(2n\pi z)}$ expanded in (even) powers of $\cos{\pi z}$, we obtain 
\be
Z_{\rm op,cl}^{(L \,\rm loops)}(q) = {1 \over \eta(q)} \: {1 \over 2^L \, (2L)!} \Big\{ \delta_{L,0} + 2 \sum_{n=L}^\infty \; T_{2n}^{(2L)}(0) \: q^{n^2/2} \Big\},
\ee
with $T_k^{(\ell)} = {\rm d}^\ell_x\,T_n(x)\big|_{x=0}$.

As particular cases, from the coefficients  (most easily derived from the Tchebychev differential equation)
\be
T_{2n}(0) = (-1)^n, \quad T_{2n}^{(2)}(0) = 4(-1)^{n+1} n^2, \quad T_{2n}^{(4)}(0) = 16 (-1)^{n}n^2(n^2-1),
\ee
we obtain the following expressions for the first three partition functions,
\bea
&& Z_{\rm op,cl}^{(\rm no\ loop)}(q) = {\theta_4 \over \eta}(q), \qquad 
Z_{\rm op,cl}^{(1\, \rm loop)}(q) = {1 \over 4\pi^2} \,{\theta''_4 \over \eta}(q), \\
\noalign{\medskip}
&& Z_{\rm op,cl}^{(2\, \rm loops)}(q) = {1 \over 48\pi^4} \, \Big[{\theta''''_4 \over \eta}(q) + 4\pi^2 \,{\theta''_4 \over \eta}(q) \Big],
\eea
where the derivatives of $\theta_4(e^{2{\rm i}\pi z}|q)$ are taken with respect to $z$, evaluated at $z=0$.


These results may be used to compute the probability distribution for the number of loops in oriented spanning webs when these are uniformly weighted. For the boundary conditions considered here, the probability that a random arrow configuration contain exactly $L$ loops when the arrows can point outside of the cylinder along one of the two edges is given by
\be
{\rm Prob}_{\rm op,cl}(L\ {\rm loops}) = {Z_{\rm op,cl}^{(L \,\rm loops)}(q) \over \sum_{L \geq 0} \: Z_{\rm op,cl}^{(L \,\rm loops)}(q)}.
\ee

The distribution generating function can be obtained from the expressions found. Setting $x = 2\cos^2{\pi z}$ and using the product representation of $\theta_3(e^{2{\rm i}\pi z}|q)$, we have
\bea
F(x) &\equiv& \sum_{L=0}^\infty \: x^L \; Z_{\rm op,cl}^{(L \,\rm loops)}(q) = q^{-1/24} \: \prod_{n=0}^\infty (1 + 2 \, q^{n+1/2} \, \cos{2\pi z} + q^{2n+1})\nonumber\\
&=& q^{-1/24} \: \prod_{n=0}^\infty \:2q^{n+1/2} \Big(\!\cosh{[(n+\textstyle{1\over 2})\pi\tau]} + x-1\Big)\,, \qquad \tau = {N \over M}.
\eea

\begin{figure}[t]
\begin{center}
\pspicture(-3,0)(2,6)
\rput(10.2,0.3){$1/\tau$}
\includegraphics[scale=.85]{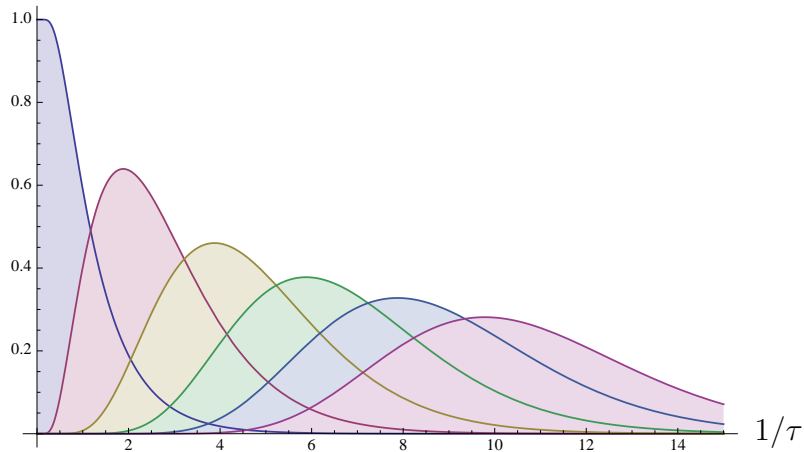}
\endpspicture
\end{center}
\caption{Plots of the probabilities ${\rm Prob}_{\rm op,cl}(L\ {\rm loops})$ as function of the inverse aspect ratio $1 \over \tau$, for $0 \leq L \leq 5$. Their generating function is given in (\ref{opcl}).}
\label{fig5}
\end{figure}

As $x=1$ corresponds to $z={1 \over 4}$, it follows that the generating function is given by
\be
\sum_{L=0}^\infty \: x^L \; {\rm Prob}_{\rm op,cl}(L\ {\rm loops}) = {F(x) \over F(1)} = \prod_{n=0}^\infty \: {\cosh{[(n+{1\over 2})\pi\tau]} + x - 1 \over \cosh{[(n+{1\over 2})\pi \tau]}}.
\label{opcl}
\ee
For fixed $L$, the probability ${\rm Prob}_{\rm op,cl}(L\ {\rm loops})$, as a function of $\tau$ has a bell-like shape, with a unique maximum around ${1 \over \tau} = 2L$. The distributions for the first values of $L$ are shown in Figure \ref{fig5}. For large $M$, small $\tau$, the average value $\la L \ra$ increases linearly with $M$.

We stress again the fact that the previous distribution refers to oriented spanning webs having a certain number of loops. One may decide to ignore the orientation and simply focus on the set of lattice bonds used by the spanning webs. All $2^L$ oriented spanning webs with $L$ loops then get all identified to a single unoriented graph, called an incompressible cycle-rooted spanning forest \cite{kenyon2011}. The number of those containing $L$ loops equal $2^{-L} Z^{(L \, \rm loops)}$ so that the generating function must be slightly reinterpreted. The resulting distribution reads
\be
\sum_{L=0}^\infty \: x^L \; {\rm Prob}^{\rm unor}_{\rm op,cl}(L\ {\rm loops}) = {F({x \over 2}) \over F({1 \over 2})} = \prod_{n\geq 0} \: {2\cosh{[(n+{1 \over 2})\pi\tau]} + x - 2 \over 2\cosh{[(n+{1 \over 2})\pi\tau]} - 1}\,. \\
\ee
The distributions in this ensemble are similar to those for oriented loops shown in Figure \ref{fig5}, the main difference being that the curves are slightly shifted to the right (except the first one, for $L=0$) and appear to have a larger dispersion.

\subsection{Open/open and closed/closed boundary conditions}

This case corresponds to an extended dimer lattice with an odd height, which we take to be $2M+1$. The relevant result from \cite{rasru} reads
\be
{\cal Z}_{2M+1,2N}(z) = {\theta_2(e^{2{\rm i}\pi z}|q) \over \eta(q)}\,, \qquad q =  \exp{(-\pi N/M)}.\ee
The usual dimer partition function ${\theta_2 \over \eta}(q)$ is recovered upon setting $z=0$ \cite{Izm03}, while the function at $z={1 \over 2}$ vanishes, as expected.

The boundary conditions are no longer the same on the odd and even sublattices. However due to the relation $Z_{\rm cl,cl}^{(L+1 \,\rm loops)} = 2\, Z_{\rm op,op}^{(L \,\rm loops)}$ in the scaling limit, it is enough to focus on one of them, say the odd sublattice, on which the two boundary conditions are closed.

Introducing the odd Tchebychev polynomials in the following identities,
\be
\hspace{-2cm}
{\cal Z}_{2M+1,2N}(z) = \sum_{L=1}^\infty \; 2^{L-1} \: Z_{\rm cl,cl}^{(L \,\rm loops)}(q) \: (\cos{\pi z})^{2L-1} = {2 \over \eta(q)} \sum_{n=1}^\infty \; \cos{(2n-1)\pi z} \: q^{(n-1/2)^2/2}\,,
\ee
yields the partition functions for the spanning webs on $\L_{\rm odd}$ with a fixed number of loops,
\be
Z_{\rm cl,cl}^{(L \,\rm loops)}(q) = {1 \over \eta(q)} \: {1 \over 2^{L-2} \, (2L-1)!} \: \sum_{n=1}^\infty \; T_{2n-1}^{(2L-1)}(0) \: q^{(n-1/2)^2/2}.
\ee
 
From $T'_{2n-1}(0) = (-1)^{n+1} (2n-1)$ and $T'''_{2n-1}(0) = (-1)^{n} (2n)(2n-1)(2n-2)$, we obtain the first two partition functions, 
\be
Z_{\rm cl,cl}^{(1\,\rm loop)}(q) = {1 \over \pi}{\theta'_1 \over \eta}(q) = {1 \over 2} \eta^2(q), \qquad Z_{\rm cl,cl}^{(2\, \rm loops)}(q) = {1 \over 12\pi^3} \,\Big\{{\theta'''_1 \over \eta}(q) + \pi^2 \, {\theta'_1 \over \eta}(q)\Big\}.
\ee


The calculation of the probability distribution for the number of loops and fully closed boundary conditions (no arrow can point outside of the cylinder), namely
\be
{\rm Prob}_{\rm cl,cl}(L\ {\rm loops}) = {Z_{\rm cl,cl}^{(L \,\rm loops)}(q) \over \sum_{L \geq 1} \: Z_{\rm cl,cl}^{(L \,\rm loops)}(q)}.
\ee
is similar to the previous. The result is
\be
\sum_{L=1}^\infty \: x^L \; {\rm Prob}_{\rm cl,cl}(L\ {\rm loops}) = x \: \prod_{n\geq 1} \: {\cosh{(\pi n \tau)} + x - 1 \over \cosh{(\pi n\tau)}}\,, \qquad \tau = {N \over M}.
\label{clcl}
\ee
The corresponding result for non-oriented spanning webs, also called incompressible cycle-rooted spanning forests, reads


\be
\sum_{L=1}^\infty \: x^L \; {\rm Prob}^{\rm unor}_{\rm cl,cl}(L\ {\rm loops}) = x \: \prod_{n\geq 1} \: {2\cosh{(\pi n \tau)} + x - 2 \over 2\cosh{(\pi n \tau)} - 1}\,, 
\label{unorclcl}
\ee
and reproduces the result obtained in \cite{kenyon2011} by different methods.


\section{Transfer matrix for spanning webs}
\label{sec6}

In the previous sections, we have computed the cylinder partition functions for spanning webs with a fixed number of loops by using the original transfer matrix proposed by Lieb, accompanied by the insertion of the defect-like operator $w^{2\cal V}$ to keep track of the loops. It is instructive to reformulate the problem by using a modified transfer matrix that, at each step of the transfer, allows to keep track of the loops. In other words, the operator $w^{2\cal V}$ should be spread out over the whole lattice instead of being concentrated on a particular row (or pair of rows). The discussion depends on the direction in which the transfer is performed.

\subsection{Transfer along perimeter (open channel)}

\begin{figure}[t]
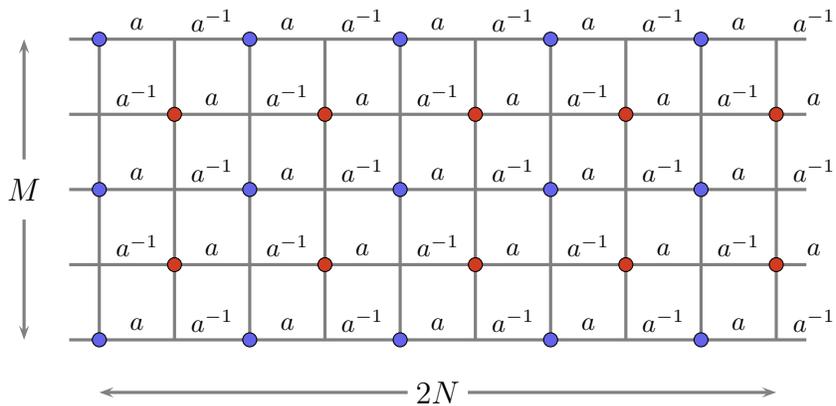

\psset{xunit=1cm}
\psset{yunit=1cm}
\psset{runit=1cm}
\begin{center}
\pspicture(-1.5,-0.5)(10,5)
\multido{\nt=0+1}{5}{\psline[linewidth=1.2pt,linecolor=gray](-0.4,\nt)(9.4,\nt)}
\multido{\nt=0+1}{10}{\psline[linewidth=1.2pt,linecolor=gray](\nt,0)(\nt,4)}
\multido{\ny=0.2+2.0}{3}{\multido{\nx=0.5+2.0}{5}{\rput(\nx,\ny){\footnotesize $a$}}}
\multido{\ny=0.25+2.00}{3}{\multido{\nx=1.5+2.0}{5}{\rput(\nx,\ny){\footnotesize $a^{-1}$}}}
\multido{\ny=1.25+2.00}{2}{\multido{\nx=0.5+2.0}{5}{\rput(\nx,\ny){\footnotesize $a^{-1}$}}}
\multido{\ny=1.2+2.0}{2}{\multido{\nx=1.5+2.0}{5}{\rput(\nx,\ny){\footnotesize $a$}}}
\multido{\ny=0+2}{3}{\multido{\nx=0+2}{5}{\pscircle[linewidth=0.4pt,fillstyle=solid,fillcolor=myblue](\nx,\ny){0.1}}}
\multido{\ny=1+2}{2}{\multido{\nx=1+2}{5}{\pscircle[linewidth=0.4pt,fillstyle=solid,fillcolor=lightred](\nx,\ny){0.1}}}
\rput(4.5,-0.7){$2N$}
\psline[linewidth=1.2pt,linecolor=gray]{->}(4.1,-0.7)(0,-0.7)
\psline[linewidth=1.2pt,linecolor=gray]{->}(4.9,-0.7)(9,-0.7)
\rput(-1,2){$M$}
\psline[linewidth=1.2pt,linecolor=gray]{->}(-1,1.6)(-1,0)
\psline[linewidth=1.2pt,linecolor=gray]{->}(-1,2.4)(-1,4)
\endpspicture
\end{center}
\caption{The figure shows the weights associated to horizontal dimers used in the text to disentangle the configurations with different number of loops.}
\label{fig6}
\end{figure}

We start with the easier case where the transfer is made in the direction of the perimeter, like in Section \ref{sec5}. Using the anticommutation of $\cal V$ and $T$, it is not difficult to see that the dimer partition function from which the partial partition functions for the spanning webs are to be extracted, can be written
\be
{\cal Z}_{M,2N} = {\rm Tr}\,(T^{2N} \, w^{2\cal V}) = {\rm Tr}\,(w^{-{{\cal V} \over N}} \, T \, w^{{\cal V}\over N} \, T)^N = {\rm Tr}\,(w^{-{{2\cal V} \over N}} \, T^2)^N.
\label{pf}
\ee
Thus the cylinder partition functions can be computed by acting alternatively with the single row modified transfer matrices $w^{-{{\cal V} \over N}} T$ and $w^{{{\cal V} \over N}} T$, or equivalently, by using the two-row modified matrix $w^{-{{2\cal V} \over N}} \, T^2$ or $T^2 \, w^{{{2\cal V} \over N}}$. It also means that the pattern of alternating weights $w$ and $w^{-1}$ located on two rows of vertical bonds, see Figure \ref{fig4}, should be copied to all vertical bonds while replacing $w$ by the new weight $a = w^{1/N}$. The result is pictured in Figure \ref{fig6}, where the cylinder has been once again rotated.

The computation of the eigenvalues of the two-row transfer matrix $T^2 \, a^{2\cal V}$ is straightforward since $T^2$ and $\cal V$ commute. A common eigenbasis has been discussed in \cite{rasru}, from which the eigenvalues are easily derived. For $M$ even for instance (we recall that $T^2 \, a^{2\cal V}$ acts on a row of $M$ sites and has dimension $2^M$), we obtain that each eigenvalue of $T^2 \, a^{2\cal V}$ has the form
\be
\lambda = \prod_{k=1}^{M/2} \: \lambda_k,
\ee
where for each $k$, $\lambda_k$ takes one of the following four values,
\be
\lambda_k = a^2, \: a^{-2}, \: [\sqrt{1+\cos^2{q_k}} + \cos{q_k}]^2, \: [\sqrt{1+\cos^2{q_k}} - \cos{q_k}]^2,
\ee
with $q_k = {\pi k \over M+1}$. The use of these eigenvalues to compute the trace in (\ref{pf}) leads directly to the results mentioned in the Section \ref{sec5}.

\subsection{Transfer along height (closed channel)}

More interesting is the case where the transfer is made in the direction of the height of the cylinder, which implies that the seeked transfer matrix will act on periodic row configurations. In conformal terms, this transfer matrix yields the conformal spectrum in the bulk (non-chiral).

According to Figure \ref{fig6}, the weights $a$ and $a^{-1}$ now refer to all the horizontal bonds, alternating both horizontally and vertically. A horizontal bond with a weight $a$ means that a dimer covering this bond gets a weight $a$. Thus the transfer matrix should not assign a uniform weight $\alpha$ to all horizontal dimers, like in Section \ref{sec3}, but alternating weights $a$ and $a^{-1}$. From the expression (\ref{tm}), it follows that the correct transfer matrix for this case has the form
\be
\widetilde T(a) = \exp{(\sum_{i=1}^{2N} \, a^{-\epsilon_i}\,\s_i^-\s_{i+1}^-)} \, \prod_{i=1}^{2N} \: \s_i^x\,, \qquad \epsilon_i = (-1)^{i},
\label{tw}
\ee
with the periodic boundary condition $\s^-_{2N+1} \equiv \s^-_1$. It assigns a weight $a$ to the first horizontal bond, a weight $a^{-1}$ to the second one, and so on. Since the weight also alternate vertically, the transfer is performed by alternatively using $\widetilde T(a)$ and $\widetilde T(a^{-1})$.

One may also define the two-row transfer matrix,
\be
\widetilde T_2(a) = \widetilde T(a^{-1}) \, \widetilde T(a) = \exp{(\sum_{i=1}^{2N} \, a^{\epsilon_i}\,\s_i^-\s_{i+1}^-)} \; \exp{(\sum_{i=1}^{2N} \, a^{-\epsilon_i}\,\s_i^+\s_{i+1}^+)},
\ee
easier to handle (and sufficient when the height $M$ is even). From what we have discussed in Section \ref{sec3} and more specifically from (\ref{str}), the dimer partition function on a cylinder with an even height $M$ is given by
\be
{\cal Z}_{M,2N} = \la\downarrow\downarrow \ldots \downarrow\!|\:\widetilde T_2(a)^{M/2}\:|\!\downarrow\downarrow \ldots \downarrow\ra.
\label{66}
\ee
It is to be noted that $\widetilde T_2(a)$ is the proper matrix to compute the dimer partition function from which we can extract the partition functions for spanning webs with a fixed number of loops. It is however not in itself a transfer matrix for spanning webs having a fixed number of loops. A notable exception is $\widetilde T_2(a={\rm e}^{{\rm i}\pi/2N})$, which corresponds to $w = a^N = {\rm i}$. As discussed in Section \ref{sec4}, this value of $w$ assigns a weight 0 to the spanning webs containing non-trivial loops, so that $\widetilde T_2(a={\rm e}^{{\rm i}\pi/2N})$ is a genuine transfer matrix to iteratively build and count spanning forests.

In the rest of this section, we report on the exact spectrum of $\widetilde T_2(a)$ and compute, in the next section, the conformal spectrum generating function in the scaling limit. We note that if $a$ is a pure phase, $\widetilde T_2(a)$ is hermitian and therefore diagonalizable. For general $a$, it is no longer hermitian, nor even normal, but is nevertheless fully diagonalizable, except at a finite number of isolated points in the complex $a$ plane, located on two circles of radii $(\sqrt{2} \pm 1)$, see below.

For $a=1$ or $w=1$, the explicit diagonalization has been carried out by Lieb \cite{lieb} using a Jordan-Wigner transformation. The same method works for a generic value of $a$, so we merely quote the results. 

The Jordan-Wigner transformation maps the spin operators $\s_i^\pm$ to fermionic creation and annihilation operators and so realizes the spin configuration space $\C^{\otimes 2N}$ as a Fock space. The even (odd) sector comprises the states of the Fock space which have an even (odd) number of fermionic excitations; in terms of spins, the even (odd) sector contains the configurations which have an even (odd) value of $\cal V$ \cite{rasru}. The splitting into sectors provides a first block-diagonalization of the transfer matrix, since the latter does not change the parity of the fermion number, 
\be
\widetilde T_2(a) = \pmatrix{\widetilde T_2^{\,\rm even} & 0 \cr 0 & \widetilde T_2^{\,\rm odd}}.
\ee
In addition, the fermions do not satisfy the same boundary condition in the two sectors, so that the two blocks are slightly different. As before, we set $w = {\rm e}^{{\rm i}\pi z}$ and $a = {\rm e}^{{\rm i}\pi z/N}$.

Carrying the Jordan-Wigner transformation, one finds that the blocks can be written in the following way,
\bea
\hspace{-1.3cm} && \widetilde T^{\rm even}_2 = \left. \bigotimes_{k=0}^{N-1} \; \Big[\exp{(2 \a_k \; \psi^{}_{k} \: \psi^{}_{2N-k-1})} \; \exp{(2 \a_k \; \psi^\dagger_{2N-k-1} \: \psi^\dagger_{k})}\Big]\right|_{\rm even}, \quad \a_k = \sin{\textstyle{\pi(z+k+1/2) \over N}},\\
\noalign{\medskip}
\hspace{-1.3cm} && \widetilde T^{\rm odd}_2 = \left. \bigotimes_{k=0}^{N-1} \; \Big[\exp{(2 \beta_k \; \psi^{}_{k} \: \psi^{}_{2N-k})} \; \exp{(2 \beta_k \; \psi^\dagger_{2N-k} \: \psi^\dagger_{k})}\Big]\right|_{\rm odd}, \qquad \beta_k = \sin{\textstyle{\pi(z+k) \over N}},
\eea
where the fermions satisfy the canonical anticommutation relations, 
\be
\{\psi^{}_i,\psi^{}_j\} = \{\psi_i^\dagger,\psi_j^\dagger\} = 0, \qquad \{\psi_i^{},\psi_j^\dagger\} = \delta_{i,j}.
\ee

Each factor in the tensor products acts in a four-dimensional Fock space, but the restriction to the even or odd sector respectively, which applies to the whole products and not to the factors separately, reduces to $2^{2N-1}$ the dimension of the two blocks $\widetilde T^{\rm even}_2$, $\widetilde T^{\rm odd}_2$. A factor in the products has the typical form given by
\be
A = \exp{(2 \gamma \; \psi_1 \: \psi_2)} \; \exp{(2 \gamma \; \psi^\dagger_2 \: \psi^\dagger_1)},
\ee
and is easily diagonalized in the following basis,
\be
\Phi_0=|0\rangle, \quad \Phi_1 = \psi^\dagger_1|0\rangle, \quad \Phi_2 = \psi^\dagger_2|0\rangle, \quad \Phi_{21} = \psi^\dagger_2\,\psi^\dagger_1|0\rangle.
\ee
The two odd states $\Phi_1$ and $\Phi_2$ are eigenstates of $A$ with eigenvalue 1. The other two eigenvectors are linear combinations of $\Phi_0$ and $\Phi_{21}$ and therefore even,
\be
v_\pm = {1 \over \sqrt{1 + \mu_\pm^2}}\;\Big\{\pm \mu_\pm \, \Phi_0 + \Phi_{21} \Big\},
\label{eig}
\ee
with eigenvalues 
\be
\lambda_\pm = \mu_\pm^2\,, \qquad \mu_\pm = \sqrt{1+\gamma^2} \pm \gamma.
\ee
We see that if $\gamma$ satisfies $\gamma^2 = -1$, $A$ is no longer diagonalizable, but contains a two-dimensional Jordan cell. For $\gamma = \a_k$ or $\beta_k$, the condition $\gamma^2=-1$ implies that $a = {\rm e}^{{\rm i}\pi z/N}$ has a complex norm equal to $(\sqrt{2} \pm 1)$. Thus $\widetilde T_2(a)$ is diagonalizable for all $a$ lying on the unit circle, as noted before.

We obtain the spectra of $\widetilde T^{\rm even}_2$ and $\widetilde T^{\rm odd}_2$, which together form that of $\widetilde T_2(a)$, as the sets
\bea
&& \lambda^{\rm even} = \prod_{k=0}^{N-1} \Big\{1 \;\; {\rm\underline {or}} \;\; 1  \;\; {\rm\underline {or}} \;\; [\sqrt{1+\a_k^2} + \a_k^{}]^2 \;\; {\rm\underline {or}} \;\; [\sqrt{1+\a_k^2} - \a_k^{}]^2\Big\}\,,\\
&& \lambda^{\rm odd} = \prod_{k=0}^{N-1} \Big\{1 \;\; {\rm\underline {or}} \;\; 1  \;\; {\rm\underline {or}} \;\; [\sqrt{1+\beta_k^2} + \beta_k^{}]^2 \;\; {\rm\underline {or}} \;\; [\sqrt{1+\beta_k^2} - \beta_k^{}]^2\Big\}\,,
\eea
where the total number of 1's in the product must be even for $\lambda^{\rm even}$ and must be odd for $\lambda^{\rm odd}$.


\section{Bulk spectrum generating functions}

The usual statement is that the spectrum of the transfer matrix on a cylinder (with periodic spatial direction) gives the full information on the spectrum of scaling operators in the bulk. This information is traditionally encoded in the torus partition function,
\be
{\cal Z} = {\rm Tr}\,\widetilde T_2^M = \sum_\lambda \: \lambda^M,
\ee
where the trace enforces periodicity in the vertical (temporal) direction. When we keep a generic value of $w$, the interpretation of ${\cal Z}$ in terms of a loop gas is however no longer clear, so we prefer to talk of ${\cal Z}$ as a bulk spectrum generating function. 

The generating function is given for arrow configurations living on an $M \times N$ grid, corresponding to a $2M \times 2N$ grid for the underlying dimer configurations. In the scaling limit, and for $w = {\rm e}^{{\rm i}\pi z}$ as before, the function ${\cal Z}$ is a function of the aspect ratio $M/N$ through the variable $q=\exp{(-2\pi M/N)}$, and a function of $z$. The value $z=0$ corresponds to dimers, whereas $z={1 \over 2}$ corresponds to spanning trees in the sense of the previous section.  

The spectrum of $\widetilde T_2(a)$ has two parts, the odd one and the even one. From the previous section, the generating function for the odd part of the spectrum reads
\be
{\cal Z}^{\rm odd}(z) = \prod_{k=0}^{N-1} \Big[1 + 1 + (\sqrt{1+\beta_k^2} + \beta^{}_k)^{2M} + (\sqrt{1+\beta_k^2} - \beta^{}_k)^{2M}\Big]\Big|_{\rm odd}\,,
\ee
where $\b_k = \sin{\pi(k+z) \over N}$, and the odd part of the product means that we do not take all terms in the product but only those terms which contain an odd number of 1's. This constraint can be taken into account by introducing a variable $y$ in the following way,
\be
{\cal Z}^{\rm odd}(z) = \prod_{k=0}^{N-1} \Big[y^{} + y^{-1} + (\sqrt{1+\beta_k^2} + \beta^{}_k)^{2M} + (\sqrt{1+\beta_k^2} - \beta^{}_k)^{2M}\Big]\,,
\ee
and keeping the odd part in $y$. The function ${\cal Z}^{\rm odd}$ can also be written as
\bea
{\cal Z}^{\rm odd} &=& \prod_{k=0}^{N-1} \Big(\sqrt{1+\beta_k^2} + \beta^{}_k\Big)^{2M} \nonumber\\
&& \hspace{3mm} \times \prod_{k=0}^{N-1} \Big\{1 + (y^{} + y^{-1})\Big(\sqrt{1+\beta_k^2} + \beta^{}_k\Big)^{-2M} + \Big(\sqrt{1+\beta_k^2} + \beta^{}_k\Big)^{-4M}\Big\}.
\eea

The first product can be computed asymptotically by using Euler-MacLaurin's formula, and yields
\be
\log{\prod_{k=0}^{N-1} \Big[\sqrt{1+\beta_k^2} + \beta^{}_k\Big]^{2M}} = {4{\rm G} \over \pi}MN + \Big[2\pi z(1-z) - {\pi \over 3}\Big]{M \over N} + \ldots
\label{fren}
\ee
up to terms which vanish in the limit $M,N \to \infty$ with $M/N$ fixed; G is the Catalan constant. 

The second product, using the following estimate,
\be
\Big(\sqrt{1+\beta_k^2} + \beta^{}_k\Big)^{-2M} \simeq \Big(1 + {\pi(k+z) \over N}\Big)^{-2M} \simeq e^{-2\pi (k+z)M/N} = q^{k+z},
\ee
as well as the identity $\a_{N-k}(z) = \a_k(-z)$, yields in the limit $N \to \infty$,
\bea
&& \hspace{-16mm}\Big[1 + (y^{} + y^{-1})\:q^z + q^{2z}\Big] \; \prod_{k=1}^\infty \Big[1 + (y^{} + y^{-1})\:q^{k+z} + q^{2(k+z)}\Big] \: \Big[1 + (y^{} + y^{-1})\:q^{k-z} + q^{2(k-z)}\Big] \nonumber\\
&=& \Big[1 + (y^{} + y^{-1})\:q^z + q^{2z}\Big] \; \prod_{k=1}^\infty \: (1 + y q^zq^k) \, (1 + y^{-1}q^{-z}q^k) \, (1 + y q^{-z} q^k) \, (1 + y^{-1}q^{z}q^k) \nonumber\\
&=& q^{z-1/6} \; {\theta_2(yq^z|q) \over \eta(q)} \: {\theta_2(y^{-1}q^z|q) \over \eta(q)}.
\eea

Combining this last result with (\ref{fren}) from which we remove the extensive term proportional to $MN$, and taking the odd part in $y$, we obtain the universal function for the odd sector as
\be
{\cal Z}^{\rm odd}(q;z) = q^{z^2} \; {\theta^2_1(q^z|q) + \theta^2_2(q^z|q) \over 2 \eta^2(q)}.
\ee

A similar calculation for the even sector yields, 
\be
{\cal Z}^{\rm even}(q;z) = q^{(z+1/2)^2} \; {\theta^2_2(q^{z+1/2}|q) - \theta^2_1(q^{z+1/2}|q) \over 2 \eta^2(q)} = q^{z^2} \; {\theta^2_3(q^z|q) + \theta^2_4(q^z|q) \over 2 \eta^2(q)},
\ee
where we have used $\theta_1(y\sqrt{q}|q) = {{\rm i} \over \sqrt{y}} q^{-1/8} \theta_4(y|q)$ and $\theta_2(y\sqrt{q}|q) = {1 \over \sqrt{y}} q^{-1/8} \theta_3(y|q)$.

Adding the odd and even parts, we obtain the full generating function for the spectrum of the transfer matrix $\widetilde T_2$ on a cylinder,
\be
{\cal Z}(q;z) = q^{z^2} \; {\theta^2_1 + \theta^2_2 + \theta^2_3 + \theta^2_4 \over 2 \eta^2}(q^z|q).
\ee

This formula reproduces the known result \cite{ferd} for the dimer model, corresponding to $z=0$,
\be
{\cal Z}_{\rm dimer}(q) = {\cal Z}(q;0) = {\theta^2_2 + \theta^2_3 + \theta^2_4 \over 2 \eta^2}(q) = \chi^{}_{(-1/8,-1/8)} + \chi^{}_{(3/8,3/8)} + \chi^{}_{\cal R}.
\ee
This function is modular invariant and a genuine partition function. It is also the torus partition function of the triplet theory \cite{gaka}, with central charge $c=-2$, and can be expressed in terms of non-chiral characters of ${\cal W}(1,2)$, as shown by the second expression in the previous equation.

The other instructive case is $z={1 \over 2}$, related to spanning trees, equivalently to the sandpile model. Interestingly, we find
\be
{\cal Z}_{\rm trees}(q) = {\cal Z}(q;{\textstyle{1 \over 2}}) = {\theta^2_2 + \theta^2_3 - \theta^2_4 \over 2 \eta^2}(q) = \chi^{}_{(-1/8,3/8)} + \chi^{}_{(3/8,-1/8)} + \chi^{}_{\cal R}.
\ee
As anticipated, it is not modular invariant since the two directions are not treated in the same way. However from its expression in terms of ${\cal W}$-characters, we see that ${\cal Z}_{\rm trees}$ appears as the $\Z_2$-twisted version of the modular invariant ${\cal Z}_{\rm dimer}$.

We have used in this section the transfer matrix $\widetilde T_2(a)$ and powers thereof, by which all horizontal bonds are assigned alternating weights $a$ and $a^{-1}$ whereas vertical bonds have a constant weight 1. We have seen in Section \ref{sec6} that this setting (in Figure \ref{fig6}) was in fact equivalent to the situation in which only two columns of horizontal bonds are assigned alternating weights $w=a^N$ and $w^{-1} = a^{-N}$ (rotating the Figure \ref{fig4}). The insertion of these weights can be viewed as a defect line propagating vertically, which is the rotated version of the seam shown in Figure \ref{fig4}, itself implemented by the operator $w^{2\cal V}$. For $z={1 \over 2}$, this operator $w^{2\cal V} = {\rm e}^{{\rm i}\pi {\cal V}}$ is a $\Z_2$ symmetry transformation since $\cal V$ takes integer values. 

Thus ${\cal Z}_{\rm trees}$ can be seen as the partition function for dimers complemented by the insertion of the $\Z_2$ defect line running vertically, whose realization in the rotated lattice is provided by ${\rm e}^{{\rm i}\pi {\cal V}}$. It gives a lattice realization of the $\Z_2$ defect in the triplet theory and explains the relation of ${\cal Z}_{\rm trees}$ to ${\cal Z}_{\rm dimer}$ \cite{spo}.

We finish this section by observing that the generating function ${\cal Z}(z;q)$ has an alternative expression in the other channel, corresponding to $\tilde q = \exp{(-2\pi N/M)}$ (exchange of $M$ and $N$, $\tau \to \tilde \tau = -1/\tau$). Indeed the theta functions satisfy the following modular transformations,
\bea
&& \theta_1(q^z|q) = {\rm i} (-{\rm i}\tau)^{-1/2} \, q^{-z^2/2} \, \theta_1(e^{2{\rm i}\pi z}|\tilde q), \\
\noalign{\medskip}
&& \theta_{k}(q^z|q) = (-{\rm i}\tau)^{-1/2} \, q^{-z^2/2} \, \theta_{6-k}(e^{2{\rm i}\pi z}|\tilde q), \qquad k=2,3,4.
\eea
Together with $\eta(q) = (-{\rm i}\tau)^{-1/2} \, \eta(\tilde q)$, they recast the generating function into 
\be
{\cal Z}(z;q) = {-\theta^2_1 + \theta^2_2 + \theta^2_3 + \theta^2_4 \over 2 \eta^2}(e^{2{\rm i}\pi z}|\tilde q).
\ee
For $z=0$, this formula explicitly shows the modular invariance of the dimer partition function.


\section{Transfer matrix for colored spanning forests}

Dimers, spanning trees, spanning webs, sandpiles and dense polymers are very closely related models. All of them, although to different extent, are believed to contain distinctive features that lead to their description in the scaling limit by logarithmic conformal field theories \cite{iprh,jpr,pera,bgpt}. The hallmark of logarithmic theories is the presence of reducible indecomposable representations in their spectrum, causing logarithms to appear in correlation functions. In particular, the Hamiltonian (or the Virasoro modes $L_0, \bar L_0$) of such theories turns out to be non-diagonalizable, namely contains Jordan cells.

For the critical dense polymer model, and the result is expected also for the infinite family of logarithmic lattice models studied in \cite{prz}, the Jordan cells are already present in the finite volume model, since the finite transfer matrix is non-diagonalizable \cite{samd}. For the other aforementioned models, it is not the case. Lieb's transfer matrix originally defined for dimers, and its generalizations described earlier in Section \ref{sec6}, turn out to be fully diagonalizable at finite volume. Whether Jordan cells emerge in the scaling limit, and how, remains an open and intriguing question. 

We will not try to answer the question in this last section. What we would like to do is to define a new transfer matrix for spanning trees, essentially based on Lieb's matrix for dimers, but dressed with new degrees of freedom, namely colors. For simplicity, we will restrict to the $M \times N$ strip, with an odd number $N$ of columns.

The idea of the construction, described below, is the following. We start with the standard Lieb transfer matrix generating fully packed dimer configurations on the strip. Then, we consider the sublattice of sites with odd horizontal and even vertical coordinates. Using the Temperley mapping, we define the spanning forest on that sublattice. Each tree in the forest has its root in the zeroth row of the full lattice and is thereby globally oriented downward. We ascribe a color to each root and allow it to propagate up the corresponding tree along the bonds oriented downward, left or right (but not upward: the principle of a transfer, as implemented by the transfer matrix, does not allow to transfer a color from a later time). The propagation of colors means a successive ascription of a color to a new site of the odd-even sublattice if it is connected with an already colored site by a bond oriented from the uncolored to colored site. Thus, the number of colors reached at a given row is a non-increasing function of the row number. This is reminiscent of what happens in the dense polymer model in the link state representation, where the number of defects is also non-increasing, causing the Jordan cell structure observed in the transfer matrix. We conjecture (and observe for small system sizes) that this property ensures a similar Jordan cell structure in the colored version of Lieb's transfer matrix.

We consider a strip of width $N$ in the infinite the square lattice, with $N$ sites in each row,
for $N$ odd. The state of site $i$ in row $m$ is described by a spin $\tau_{i,m} = \,\uparrow$ or $\downarrow$, and, in addition, each odd site in a row has a color $k$, between 1 and $N+1 \over 2$, or is uncolored, in which case we set $k=0$. Even sites are described by spins only. In the following, we set $K \equiv {N+1 \over 2}$.

As above, we consider the up and down spins as the canonical base elements  $\uparrow \,= \scriptsize (\matrix{1 \cr 0})$ and $\downarrow \,= \scriptsize (\matrix{0 \cr 1})$, so that a row configuration of spins is an element of $(\C^2)^{\otimes N}$. Along with the standard Pauli matrices $\sigma_i^{-}$, $\sigma_i^{+}$, $\sigma_i^{x}$, we will use the projectors
\begin{equation}
n_i^{\uparrow} = \s_i^+\s_i^- = \pmatrix{1 & 0 \cr 0 & 0},\quad
n_i^{\downarrow} = \s_i^-\s_i^+ =  \pmatrix{0 & 0 \cr 0 & 1}.
\end{equation}

For the $(K+1)$-dimensional vector denoting the color $0 \leq k \leq K$ at an odd site $i$, we set $|k\ra_{i} = (0, \dots, 0, 1,0, \dots, 0)^t$ with a 1 in position $k+1$; the vector $(1,0 \dots,0)^t$ is used for the case when no color is ascribed to the odd site. In addition, we define the projector of any color state at site $i$ onto the $\ell$-th colored state $|\ell\ra_i, \; \ell=0,1,\dots,K$, by
\begin{equation}
(P_\ell)_i = \sum_{k=0}^K \: (|\ell\ra \la k|)_i,
\end{equation}
and the indicator of a (true) color at a site by
\be
C_i = \sum_{k=1}^K \: (|k\ra \la k|)_i = \pmatrix{0&0&0&\dots&0 \cr
0&1&0&\dots&0 \cr \vdots&&&&\vdots \cr 0&0&0&\dots&1}.
\ee

Thus, the full, color and spin, configuration of row $m$ row can be written as
\be
S_m=\bigotimes_{i=1}^{K} \: |k_i\rangle_{(2i-1),m} \; \bigotimes_{j=1}^N \: \tau_{j,m} \quad \in (\C^{K+1})^{\otimes K} \otimes (\C^2)^{\otimes N}.
\ee
We specify the state of the first row, the in-state, to be
\be
S_1=\bigotimes_{i=1}^{K}|i\rangle_{(2i-1),1}\bigotimes_{j=1}^N\tau_{j,1}, \quad
\mbox{with}\; \tau_{j,1}=\uparrow \; \mbox{for all}\; j
\ee
in order to ensure the maximum number $K=(N+1)/2$ of colors at the bottom of the forest. 

The transfer matrix acts both on the spin content of a configuration and on the color variables to ensure the propagation of the colors. The transfer on the spin part is effected by the same matrix $T$ introduced in section 3, Eq. (\ref{tm}). The transfer of the color labels, which depends on whether it acts on an odd row or on an even row, is defined as follows. 

The following operators $U_{2i-1}$ are responsible for the transfer of the colors from the odd sites  of an odd row $m$ up to the next even row:
\begin{equation}
U_{2i-1} = {\Bbb I}_{2i-1} \otimes n^{\uparrow}_{2i-1} + (P_0)_{2i-1} \otimes n^{\downarrow}_{2i-1}\,,
\label{U}
\end{equation}
where the two factors in the tensor products refer respectively to the color variable and to the spin variable at site $2i-1$. A spin up means the presence of a vertical dimer connecting the sites $(2i-1,m)$ and $(2i-1,m+1)$, and therefore the presence of a bond in the tree oriented downward. This bond thus conducts the color from site $(2i-1,m)$ to site $(2i-1,m+1)$ of the next upper row; this action is described by the first term in $U_{2i-1}$. Likewise a spin down at site $2i-1$ implies the absence of a bond in the tree pointing downward from the site $(2i-1,m+1)$, so that the propagation of color is forbidden: the site $(2i-1,m+1)$ becomes uncolored after the action of the second term of $U_{2i-1}$. The operators $U_{2i-1}$ implement the vertical transfer of the colors from row $m$ to row $m+1$ but their action depends on the spins on row $m$. Therefore they should act before the spin variables themselves are transferred by $T$. It follows that 
\be
\widetilde{T}= T \: \Big[\prod_{i=1}^{K} \, U_{2i-1}\Big]
\label{tildeT}
\ee
ensures the vertical transfer of both the spin and the color variables from an odd row to the next row up.

After the transfer by $\widetilde T$, we need to worry about the possible transfer of colors horizontally in the even row $m+1$. A transfer of color between $(2i-1,m+1)$ and $(2i+1,m+1)$ is possible if and only if both sites $(2i,m+1)$ and $(2i,m)$ have spins down, since this ensures the presence of a horizontal dimer either connecting sites $(2i-1,m+1)$ and $(2i,m+1)$ (case 1), or $(2i,m+1)$ and $(2i+1,m+1)$ (case 2). In case 1, the tree has a bond pointing from $(2i-1,m+1)$ to the right so that the transfer of color goes in the opposite direction, from right to left, thus from $(2i+1,m+1)$ to $(2i-1,m+1)$. It is the opposite in case 2, for which the color of $(2i-1,m+1)$ is transferred to $(2i+1,m+1)$.

The spin at $(2i,m)$ can be probed by applying $n_{2i}^{\downarrow}$ before the action of $\widetilde T$. If indeed the spin at $(2i,m)$ is down (the projector $n_{2i}^{\downarrow}$ placed before $\widetilde T$ does not give zero), one acts with the following two commuting operators
\bea
R_{2i+1,2i-1} &=& \sum_{\ell=1}^K \: (P_\ell)_{2i-1}\,(|\ell\ra\la\ell|)_{2i+1}\,n_{2i}^{\downarrow} +
(|0\ra\la 0|)_{2i+1} \, n_{2i}^{\downarrow} + n_{2i}^{\uparrow}\,, \\
L_{2i-1,2i+1} &=& \sum_{\ell=1}^K \: (P_\ell)_{2i+1}\,(|\ell\ra\la\ell|)_{2i-1}\,n_{2i}^{\downarrow} +
(|0\ra\la 0|)_{2i-1} \, n_{2i}^{\downarrow} + n_{2i}^{\uparrow}\,.
\eea
In the first two terms of both operators, the projector $n_{2i}^{\downarrow}$ tests the spin at $(2i,m+1)$. If it is down, the first term of $R_{2i+1,2i-1}$ transfers a non-zero color from $(2i+1,m+1)$ to $(2i-1,m+1)$ (case 1), whereas that of $L_{2i-1,2i+1}$ transfers a non-zero color in the opposite direction (case 2). For each operator, the second term ensures that a zero color is not transferred, while the third term makes sure that no transfer of color takes place when site $(2i,m+1)$ has a spin up.

It should be noted that pairs of operators $L_{2i-1,2i+1}, L_{2j-1,2j+1}$ having one index in common do not commute, and the same is true of the $R$ operators. In order to correctly propagate the colors in the two directions, the $R$ operators must be applied by starting from the rightmost site, and ending at the leftmost site, i.e. in the order $R_{3,1}R_{5,3} \ldots R_{N,N-2}$; the $L$ operators must be applied in the inverse order, as $L_{N-2,N} \ldots L_{3,5}L_{1,3}$.



Putting all together, we obtain the transfer matrix $T^{odd}$ acting on odd row configurations and transferring colors upward, and leftward or rightward in the next even row, can be written as
\be
\hspace{-1cm}
T^{\rm odd} = \sum_{\{\theta_{2k} = 0,1\}} \Big[\prod_{i=1}^{K-1} \: R^{\theta_{2i}}_{2i+1,2i-1} \Big]
\Big[\prod_{i=K-1}^1 \: L^{\theta_{2i}}_{2i-1,2i+1} \Big]\;
\widetilde{T}\; \Big[\prod_{i=1}^{K-1}\:[\theta^{}_{2i} n_{2i}^{\downarrow}+ (1-\theta^{}_{2i})n_{2i}^{\uparrow}]\Big] ,
\label{Todd}
\ee
where the sum over $\{\theta_{2k} = 0,1\}$ denotes a sum over $\theta_{2}, \theta_{4}, \dots, \theta_{N-1}$, each taking the values $0,1$. By convention, we set $R^0 = L^0 = {\Bbb I}$.


The transfer matrix acting on the configuration of an even row does not transfer colors horizontally and has a much simpler form,
\begin{equation}
T^{\rm even} = \Big[\prod_{i=1}^{K} \: U_{2i-1}\Big] \; T.
\label{Teven}
\end{equation}
Horizontal transfer of colors is not needed, because it would refer to horizontal bonds in the tree living on the other sublattice (even horizontal and odd vertical coordinates), which we are not interested in. For the same reason, the order of the spin transfer matrix $T$ and the operators $U$ is reversed. 

\begin{figure}[t]
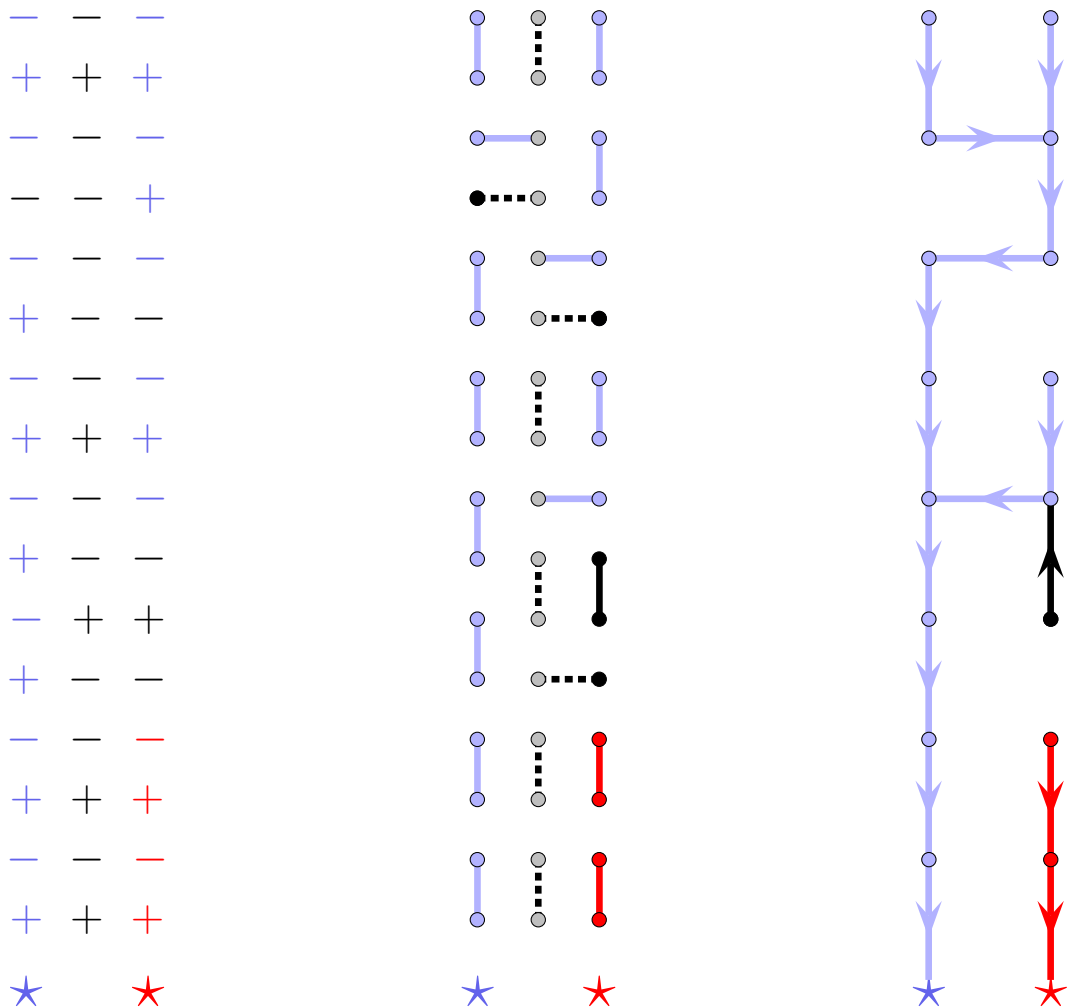

\psset{xunit=0.6cm}
\psset{yunit=0.8cm}
\psset{runit=0.6cm}
\def\pn{\Large \black $+$}
\def\pb{\Large \textcolor{myblue}{$+$}}
\def\pr{\Large \red $+$}
\def\mn{\Large \black $-$}
\def\mb{\Large \textcolor{myblue}{$-$}}
\def\mr{\Large \red $-$}
\def\br{\psline[linewidth=2.5pt,linecolor=midblue](0,0)(1.35,0)}
\def\bu{\psline[linewidth=2.5pt,linecolor=midblue](0,0)(0,1)}
\def\ru{\psline[linewidth=2.5pt,linecolor=red](0,0)(0,1)}
\def\nu{\psline[linewidth=2.5pt,linecolor=black](0,0)(0,1)}
\def\ndr{\psline[linewidth=2.5pt,linestyle=dashed,dash=3pt 2pt,linecolor=black](0,0)(1.35,0)}
\def\ndu{\psline[linewidth=2.5pt,linestyle=dashed,dash=3pt 2pt,linecolor=black](0,0)(0,1)}
\pspicture(-5,-1.8)(10,15)
\rput(0,0){
\rput(-1.35,-1.2){\huge\textcolor{myblue}{$\star$}}
\rput(1.35,-1.2){\huge\red $\star$}
\rput(0,0){\pb \ \ \pn\ \ \pr}
\rput(0,1){\mb \ \ \mn\ \ \mr}
\rput(0,2){\pb \ \ \pn\ \ \pr}
\rput(0,3){\mb \ \ \mn\ \ \mr}
\rput(0,4){\pb \ \ \mn\ \ \mn}
\rput(0,5){\mb \ \ \pn\ \ \pn}
\rput(0,6){\pb \ \ \mn\ \ \mn}
\rput(0,7){\mb \ \ \mn\ \ \mb}
\rput(0,8){\pb \ \ \pn\ \ \pb}
\rput(0,9){\mb \ \ \mn\ \ \mb}
\rput(0,10){\pb \ \ \mn\ \ \mn}
\rput(0,11){\mb \ \ \mn\ \ \mb}
\rput(0,12){\mn \ \ \mn\ \ \pb}
\rput(0,13){\mb \ \ \mn\ \ \mb}
\rput(0,14){\pb \ \ \pn\ \ \pb}
\rput(0,15){\mb \ \ \mn\ \ \mb}
}
\rput(10,0){
\rput(-1.35,-1.2){\huge\textcolor{myblue}{$\star$}}
\rput(1.35,-1.2){\huge\red $\star$}
\rput(-1.35,13){\br}
\rput(0,11){\br}
\rput(0,7){\br}
\rput(-1.35,0){\bu}
\rput(-1.35,2){\bu}
\rput(-1.35,4){\bu}
\rput(-1.35,6){\bu}
\rput(-1.35,8){\bu}
\rput(-1.35,10){\bu}
\rput(-1.35,14){\bu}
\rput(1.35,8){\bu}
\rput(1.35,12){\bu}
\rput(1.35,14){\bu}
\rput(1.35,0){\ru}
\rput(1.35,2){\ru}
\rput(1.35,5){\nu}
\rput(0,0){\ndu}
\rput(0,2){\ndu}
\rput(0,5){\ndu}
\rput(0,8){\ndu}
\rput(0,14){\ndu}
\rput(-1.35,12){\ndr}
\rput(0,4){\ndr}
\rput(0,10){\ndr}
\multido{\nt=0+1}{16}{\pscircle[linewidth=0.4pt,fillstyle=solid,fillcolor=midblue](-1.35,\nt){0.17}}
\multido{\nt=0+1}{16}{\pscircle[linewidth=0.4pt,fillstyle=solid,fillcolor=mygrey](0,\nt){0.17}}
\multido{\nt=7+1}{9}{\pscircle[linewidth=0.4pt,fillstyle=solid,fillcolor=midblue](1.35,\nt){0.17}}
\pscircle[linewidth=0.4pt,fillstyle=solid,fillcolor=black](-1.35,12){0.17}
\pscircle[linewidth=0.4pt,fillstyle=solid,fillcolor=black](1.35,10){0.17}
\multido{\nt=4+1}{3}{\pscircle[linewidth=0.4pt,fillstyle=solid,fillcolor=black](1.35,\nt){0.17}}
\multido{\nt=0+1}{4}{\pscircle[linewidth=0.4pt,fillstyle=solid,fillcolor=red](1.35,\nt){0.17}}
}
\rput(20,0){
\def\fbr{\psline[linewidth=2.5pt,linecolor=midblue,arrowsize=5pt 2]{->}(0,0)(1.65,0)}
\def\fbl{\psline[linewidth=2.5pt,linecolor=midblue,arrowsize=5pt 2]{->}(1.35,0)(-0.3,0)}
\def\fbd{\psline[linewidth=2.5pt,linecolor=midblue,arrowsize=5pt 2]{->}(0,1)(0,-0.3)}
\def\frd{\psline[linewidth=2.5pt,linecolor=red,arrowsize=5pt 2]{->}(0,1)(0,-0.3)}
\def\fnu{\psline[linewidth=2.5pt,linecolor=black,arrowsize=5pt 2]{->}(0,0)(0,1.3)}
\rput(-1.35,-1.2){\huge\textcolor{myblue}{$\star$}}
\rput(1.35,-1.2){\huge\red $\star$}
\rput(-1.35,13){\fbr}
\rput(0,13){\br}
\rput(0,11){\fbl}
\rput(-1.35,11){\br}
\rput(0,7){\fbl}
\rput(-1.35,7){\br}
\rput(-1.35,-1){\bu}
\rput(-1.35,1){\bu}
\rput(-1.35,3){\bu}
\rput(-1.35,5){\bu}
\rput(-1.35,7){\bu}
\rput(-1.35,9){\bu}
\rput(-1.35,13){\bu}
\rput(-1.35,0){\fbd}
\rput(-1.35,2){\fbd}
\rput(-1.35,4){\fbd}
\rput(-1.35,6){\fbd}
\rput(-1.35,8){\fbd}
\rput(-1.35,10){\fbd}
\rput(-1.35,14){\fbd}
\rput(1.35,14){\fbd}
\rput(1.35,12){\fbd}
\rput(1.35,8){\fbd}
\rput(1.35,7){\bu}
\rput(1.35,11){\bu}
\rput(1.35,13){\bu}
\rput(1.35,1){\ru}
\rput(1.35,-1){\ru}
\rput(1.35,6){\nu}
\rput(1.35,5){\fnu}
\rput(1.35,0){\frd}
\rput(1.35,2){\frd}
\multido{\nt=1+2}{8}{\pscircle[linewidth=0.4pt,fillstyle=solid,fillcolor=midblue](-1.35,\nt){0.17}}
\multido{\nt=5+2}{6}{\pscircle[linewidth=0.4pt,fillstyle=solid,fillcolor=midblue](1.35,\nt){0.17}}
\multido{\nt=1+2}{2}{\pscircle[linewidth=0.4pt,fillstyle=solid,fillcolor=red](1.35,\nt){0.17}}
\pscircle[linewidth=0.4pt,fillstyle=solid,fillcolor=black](1.35,5){0.17}
}
\endpspicture
\caption{Illustration of the propagation of colors in a spanning forest, defined on a vertical strip containing 3 columns. Two roots, shown by stars, are placed on the zeroth row (the bottom one) below the odd columns (the first and third). The roots carry an additional color variable, here blue and red. The sites on the second column have no color label (in gray), while the sites of the first and third columns with a zero color label are represented in black. The three panels illustrate the bijective mappings between configurations of (a) spins ($+$ is for ``up'', $-$ for ``down'') and colors, (b) colored dimers, and (c) colored forests. Tree edges oriented upward are uncolored (in black).}
\label{fig7}
\end{figure}

We thus obtain the double row transfer matrix as
\be
{\cal T} = T^{\rm even} \, T^{\rm odd},
\ee
producing spin and color configurations on odd rows. The action of the colored transfer matrices $T^{\rm odd}$ and $T^{\rm even}$ is illustrated in Fig. \ref{fig7}.

A main peculiarity of the proposed colored transfer matrix is its (partial) ability to detect connectivity between different branches of the spanning forest. 
Lieb's pure spin transfer matrix is sufficient to reproduce all possible dimer and spanning forest configurations but obviously loses the information on connectivity. As connectivity may be one of the non-local observables forcing a description by a logarithmic conformal theory, and Jordan cells in the conformal representations, we are led to suspect that the colored transfer matrix might have Jordan cells. 

Checking this out is not easy. Because of the operators responsible for the transfer of color labels, the transfer matrix is not easy to handle analytically. One can perform a numerical investigation, but its size $2^N(K+1)^K$ grows very fast: the full matrix, equal to the total number of color-spin configurations, has dimension $72, 2\,048$ and $80\,000$ for $N=3,5,7$ ! Not all configurations have to be taken in account however. For instance, the natural order of the non-zero color labels along the rows, $1 \leq k \leq K$, must be preserved by the transfer, since we start with the ordered color configuration $(1,2,\ldots,K)$ on row 1 (the colored branches cannot cross each other). More generally, the initial configuration $S_1$ gives rise, under transfer, to a limited number of subsequent configurations above it, which are the only ones worth keeping. This usually leads to a drastic drop in the number of relevant configurations. For $N=3$ and 5, the results are as follows.

For $N=3$, the full matrix $\cal T$ is 72-dimensional but its image on the initial state $S_1$ is only 7-dimensional. The corresponding subspace is generated by the following seven states (as in Fig. \ref{fig7}, the ``$+$'' and ``$-$'' denote up and down spins respectively, while the numbers refer to the colors):
\bea
1 &=& |\!+1;+;+\,1\ra, \quad 2 = |\!+1;+;+\,2\ra \hbox{\ \ (the in-state)}, \quad 3 = |\!+1;-;-\,0\ra,\\
4 &=& |\!+2;+;+\,2\ra, \quad 5 = |\!+2;-;-\,0\ra, \quad 6 = |\!-0;-;+\,1\ra,\\
7 &=& |\!-0;-;+\,2\ra.
\eea
The restriction of $\cal T$ to this invariant subspace reads explicitly
\be
M = {\cal T}^{\rm restr} = \pmatrix{
1 & 0 & 1 & 0 & 0 & 1 & 0\cr
0 & 1 & 0 & 0 & 0 & 0 & 0\cr
1 & 1 & 2 & 0 & 0 & 1 & 0\cr
0 & 0 & 0 & 1 & 1 & 0 & 1\cr
0 & 0 & 0 & 1 & 2 & 0 & 1\cr
1 & 0 & 1 & 0 & 0 & 2 & 0\cr
0 & 1 & 0 & 1 & 1 & 0 & 2}.
\ee
Its characteristic polynomial reads $(z-1)^3 \, (z^2 - 4z + 1)^2 = 0$, showing that the matrix has three different eigenvalues, $\lambda=1$, degenerate three times, and $\lambda = 2 \pm \sqrt{3}$, each twice degenerate. One can check that the equation $(z-1) \, (z^2 - 4z + 1) = 0$ is not satisfied by $M$, whereas the following is, $(M - {\Bbb I})^2 \,(M^2 - 4M + {\Bbb I}) = 0$. 
It implies that $M$ is non-diagonalizable, and possesses one Jordan cell of size 2, associated with the eigenvalue 1. 

The situation is similar for the next case $N=5$. As mentioned above, the full $\cal T$ has dimension $2\,048$, but its restriction to the image of the initial state $S_1$ has dimension 95. By looking at its characteristic polynomial as before, one finds that it has again a single Jordan block of size 2, associated with the eigenvalue 1 (which is degenerate 27 times).

We have not been able to go much further, and the only few general observations about the restriction ${\cal T}^{\rm restr}$ can be summarized as follows. It is not difficult to see that the state $S_1$ is a left eigenvector of $\cal T$ with eigenvalue 1, $S_1 {\cal T} = S_1$. In fact the eigenvalue 1 is always highly degenerate, since the states with all spins up and any sequence of non-zero colors such that two consecutive colors are not equal, are all left eigenvectors of $\cal T$ with eigenvalue 1, making it degenerate at least $K(K-1)^{K-1}$ times. In addition we have observed that $S_1$ is orthogonal to all right eigenvectors, suggesting that $S_1$ is part of a right generalized eigenvector of eigenvalue 1. 

All this is too preliminary to lead to any definite conclusion. However the fact that this new transfer matrix carries non-local degrees of freedom may open a new perspective.


\ack

This work was supported by the RFBR grant No. 12-01-00242a, the
Heisenberg-Landau program and the Belgian Interuniversity Attraction Poles Program P7/18 through the network DYGEST (Dynamical, Geometry and Statistical Physics). PR is Senior Research Associate of the Belgian Fonds de la Recherche Scientifique - FNRS; he is grateful to Alexi Morin-Duchesne for valuable comments on the manuscript.


\end{document}